\documentclass[
reprint,
superscriptaddress,
nobibnotes,
 amsmath,amssymb,amsfonts,
 aps,
prl,
floatfix,
]{revtex4-1}

\usepackage{dcolumn}
\usepackage{bm}

\usepackage[english]{babel}    
\usepackage{amsfonts}
\usepackage{amsmath,amssymb}   
\usepackage{amsfonts}
\usepackage{bm}
\usepackage[T1]{fontenc}       
\usepackage{array}
\usepackage{float}             
\usepackage[titletoc,title]{appendix} 
\usepackage{graphicx}          
\usepackage{graphics}
\usepackage{hyperref}
\graphicspath{{figs/}}  

\begin{document}

\title{Quantum electrodynamics in photonic crystals and controllability of ionization energy of atoms}

\author{Renat Kh. Gainutdinov}
 \email{Renat.Gainutdinov@kpfu.ru.; Also at Institute of Physics, Kazan Federal University, 16a Kremlevskaya St., Kazan 420008,
Russian Federation.}
\affiliation{Kazan Federal University, Kazan, 420008, Russia}
\affiliation{Tatarstan Academy of Sciences, Kazan, 420111, Russia}

\author{Adel I. Garifullin}
\affiliation{Kazan Federal University, Kazan, 420008, Russia}

\author{Marat A. Khamadeev}
\affiliation{Kazan Federal University, Kazan, 420008, Russia}
\affiliation{Tatarstan Academy of Sciences, Kazan, 420111, Russia}

\author{Myakzyum Kh. Salakhov}
\affiliation{Tatarstan Academy of Sciences, Kazan, 420111, Russia}
\affiliation{Kazan Federal University, Kazan, 420008, Russia}

\begin{abstract}{}
The periodic changes in the physical and chemical properties of the chemical elements are caused by the periodic change of the ionization
energies, which are constant for each element that manifested in the Periodic Table. However, as has been recently shown the modification of
the electromagnetic field in the photonic crystals gives rise to the modification of the electron electromagnetic mass. We show that the
effect can significantly change the ionization energy of atoms placed in voids of photonic crystals consisting of metamaterials with a highly
tunable refractive index and voids. The controllability of these materials gives rise to the controllability of the ionization energies over
a wide range.
\end{abstract}

\maketitle

\section{Introduction}

Since the experiments by Lamb and Retherford \cite{LambRetherford1947} quantum electrodynamics (QED) is one of the most precisely tested theories
of modern physics. We should mention, for example, cavity QED, in which the interactions between atoms and a single electromagnetic
mode of a high-Q cavity are studied \cite{Haroche2006,Vasco2019}; waveguide QED, where atoms, both natural and artificial, couple to various types
of one-dimensional waveguides and interact with a continuum of propagating photonic modes \cite{Mirhosseini2018}, and circuit QED, where the
superconducting artificial atoms can be coupled to quantized microwave fields in the transmission-line or 3D resonators \cite{Gu2017}.

The study of photonic crystals (PCs) is another example where QED plays an important role. PCs are artificial materials in which a periodic spatial
modulation of the refractive index leads to a gapped dispersion relation. This feature has many potential technological applications
\cite{Yablonovitch1987,John1987,John1990quantum,John1991,Quang1997coherent,Zhu1997quantum,Bay1997atom,Busch2000radiating,Lopez2003,Joannopoulos2008,
Soukoulis2012photonic,Wierer2009,Aguirre2010tunable,Huang2011dirac,Gainutdinov2012,Von2013bottom,Berman2018,Fenzl2014photonic,Goban2014atom,Segal2015controlling,
Jing2016,Ouchani2018,Hou2018,Ghasemi2019graphene,Abadla2020,Moradi2021}. From the very beginning, PCs have received a lot of attention from QED
in the investigations concerned with many interesting novel effects that do not manifest in free space, such as strong emitter-photon coupling,
coherent control of spontaneous emission, modification of Lamb shift, and others \cite{Yablonovitch1987,John1987,John1990quantum,John1991,Quang1997coherent,
Gainutdinov2012,Zhu2012highly,Liu2010observation,Roy2010coherent,Vats2002theory,Li2001optical,Wang2004giant,Entezar2009,Mirza2017,
Gainutdinov2018,Dey2019controlling,Stewart2020}.
These applications and effects are mainly based on the photonic band gap effect. However, the band gap is not only an effect caused by the
periodic change in the refractive index of PCs. In \cite{Gainutdinov2012} it is shown that the modification of the interaction of an electron
trapped in air voids of a PC with its own radiation field results in the change in its electromagnetic mass.
Being anisotropic, this correction depends on the electron state and is several orders larger than the atomic Lamb shift. In this paper, we
investigate the effect of the change of the electromagnetic mass and ionization energy of atoms placed in the PC medium.

The periodic changes in physical and chemical properties of elements is caused by the periodic change in ionization energy being the
minimal energy required to remove the outermost electron from an atom with the atomic number \textit{Z}. The ionization energy of each element is
constant and this manifests itself in the Periodic Table. However, as we show in this paper, the ionization energies can be dramatically changed,
when atoms are placed in air voids of a PC. Recently, great progress has been achieved in the design of metamaterials with unnaturally highly
tunable refractive indices (see \cite{Lee2015,Chung2016,Kim2016,Kim2018}). The effect under study is also controllable. This allows one to come beyond
the limitations put on by the Periodic Table on physical and chemical processes, and can open up new horizons for the synthesis of exceptional chemical
compounds that could be used in pharmaceutical and other medical-related activities.

Most explicitly this effect manifests itself when a PC consists of regularly arranged voids in a dielectric with a highly tunable refractive index.
The typical size of the voids is $10^3$ times larger than the Bohr radius. This means that an atom in a void behaves itself as a free one. Only the
interaction of the atom with its own radiation field is modified in the PC medium.

The physical mass $m_e$ of the electron is a sum of its bare mass $m_0$ and electromagnetic mass $m_{em}$:
\begin{equation}
m_e = m_0 + m_{em}. \notag \\
\end{equation}
However, this self-energy cannot be calculated because of the non-renormalizable ultraviolet divergences. The problem is solved by
including the electromagnetic mass into the physical mass $m_e$ being the only observable.
This is a part of the renormalization procedure being the cornerstone of QED. The above may be a reason why for a long time, since the pioneering
works of Yablonovitch \cite{Yablonovitch1987} and John \cite{John1987}, no attention had been paid to the fact that the modification of the
electromagnetic field in the PC medium results in the correction ${\delta m}_{pc}$ to the electron self-energy that cannot be hidden in the electron
physical mass. In contrast to the electromagnetic mass of an electron in vacuum, the PC correction to this mass is an observable. Actually, in this
case we deal with a fundamental quantum electrodynamic effect that manifests itself only in artificial media, such as PCs. For the first time the
electromagnetic mass comes into play in describing physical processes. It is essential that the PC correction to the electron electromagnetic mass
depends on both the electron state of the atom and the anisotropy of the electromagnetic field in the air voids of a PC. This correction is an observable
and is described by an operator. The interaction described by this operator can be so strong as to be comparable to the interaction of the valence
electrons with the atomic nucleus and significantly affects the ionization energies of atoms in the PC medium.

In Sec. 2, we show that the axial symmetry associated with a direction of refractive index modulation of a one-dimensional PC
gives rise to anisotropy of the electron self-energy interaction and, consequently, to anisotropy of the electron mass and optical spectra
of atoms placed in air voids of the PC medium. Section 3 is devoted to deriving, explaining, and estimating the ionization energy
correction of atoms in a one-dimensional PC. Furthermore, we consider the optical properties of a one-dimensional PC
based on high-index metamaterials, using for this the theory of optical effective media with independent control of permittivity and permeability,
and discuss how the controllability of these materials implies the controllability of the ionization energies of hydrogen atoms and alkali metals
over a wide range. In Sec. 4, we summarize the results of our research.

\section{Anisotropy of the electron mass in the photonic crystal medium}\label{S2}

The electromagnetic mass of an electron in vacuum is generated by its interaction with its own electromagnetic field. One of the processes
that gives the contributions to the electron self-energy is the emission and then absorption of a photon. In the Coulomb gauge, this process is
divided into two parts. The process of the interaction of the electron with its own Coulomb field, which leads to the change in the
electron mass, and the process of its interaction with its own transverse field gives rise to the correction of its kinetic energy (here and
below, we use the natural system of units, in which $\hbar = c = 1$):
\begin{equation}\label{eqn2}
\Delta {E_e}\left( {\widehat{\bf{p}}} \right) = \Delta {m_{em}} + \frac{{\Delta m_{em}^{'}}}{{2m_e^2}}{{\widehat{\bf{p}}}^2} + O{\left( {\left|
{\widehat{\bf{p}}} \right|}
\right)^3} + O\left( {\frac{{\Delta {m_{em}}}}{{{m_e}}}} \right),
\end{equation}
where $\widehat{\bf{p}}$ is the operator of the electron momentum and
\begin{equation}\label{eqn3}
\Delta m_{em}^{'} = \frac{\alpha }{{{\pi ^2}}}\int\limits_{0}^{k_0} {\frac{{{d^3}{\bf{k}}}}{{2{{\bf{k}}^2}}}} \sum\limits_{\lambda  = 1}^2 {{{\left|
{{\widehat{{\bf{I}}}_{\bf{p}}} \cdot {{\bm{\varepsilon }}_\lambda }({\bf{k}})} \right|}^2}} = \frac{{{4\alpha }}}{{{3{{\pi }}}}}{k_0}
\end{equation}
with ${{\widehat{\bf{I}}}_{\bf{{p}}}} = \frac{{\widehat{\bf{p}}}}{\left| {\widehat{\bf{p}}} \right|}$  being the operator of the direction of the
electron momentum, $\bm{\varepsilon}_\lambda  ({\bf{k}})$ denotes the unit vector of the field polarization ($\lambda$) in free space, and $\alpha$
is the fine-structure constant. The correction $\Delta m_{em}^{'}$ to the electron mass is determined by a divergent integral, unless a cutoff $k_0$ is
introduced \cite{Cohen1998atom}. This electron mass correction $\Delta m_{em}^{'}$ which appears in the expression for the correction to the kinetic
energy of the electron, must coincide with the mass correction $\Delta m_{em}$, being the result of the interaction of the electron with its own Coulomb field. This follows from the
relativistic energy-momentum relationship for an electron in vacuum,
\begin{equation}
E^2 - {\textbf{p}}^2 = m_e^2 \notag \\
\end{equation}
with $E^2 - {\textbf{p}}^2$ being a Lorentz invariant.
Indeed, the mass correction $\Delta m_{em}$ describing the one-photon Coulomb self-energy with the same cutoff is given by equation
\begin{equation}\label{eqn4}
\Delta m_{em} = \frac{{{4\alpha }}}{{{3{{\pi }}}}}{k_0}
\end{equation}
(see \cite{Cohen1998atom}). Thus, $\Delta m_{em}$, defined in Equation~(\ref{eqn4}), is a correction to the electron mass from the self-energy
processes associated with the emission and absorption of virtual photons. In the case where the electron interacts with the PC electromagnetic field,
the photons are replaced with the Bloch photons. Because of the spatial modulation of the refractive index, the eigenstates of photons in a PC
differ significantly from those in vacuum or uniform media. The solution of quantum-field Maxwell's equations yields photon states having the Bloch
structure similar to that of the states of electrons in ordinary crystals \cite{Ashcroft1976}. These vectors are the eigenvectors corresponding to
the energy $\omega_n$ and momenta $\bf{k}+\bf{G}$, where $n$ is a band index; the value of $\bf{k}$ is limited by the first Brillouin zone (FBZ) and
$\bf{G}$ is the reciprocal lattice vector of the PC (${\bf{G}} = N_1 {\bf{b}}_1  + N_2 {\bf{b}}_2 + N_3 {\bf{b}}_3$, where ${\bf{b}}_i$ are
primitive basis vectors of a reciprocal lattice). Due to the translation symmetry, the eigenfrequencies of the structure are usually
computed within the first Brillouin zone. By introducing the operators ${\hat a_{{\bf{k}}n}^\dag }$ and ${{{\hat a}_{{\bf{k}}n}}}$ that
describe the creation and annihilation of the photon in the state $\left| {{\bf{k}}n} \right\rangle$ ($\hat a_{{\bf{k}}n}^\dag
\left| 0 \right\rangle  = \left| {{\bf{k}}n} \right\rangle$ and ${{\hat a}_{{\bf{k}}n}}\left| {{\bf{k}}n} \right\rangle  = \left| 0
\right\rangle$), one can construct a modified "free" Hamiltonian $\widehat{H}_0 = \sum\limits_{{\bf{k}}n} {{\omega _{{\bf{k}}n}}} \hat
a_{{\bf{k}}n}^\dag \hat a_{{\bf{k}}n}^{}$ and quantized vector potential:
\begin{equation}\label{eqn15}
\begin{split}
{\widehat{\bf{A}}_{pc}}({\bf{r}},t) = \sum\limits_{{\bf{k}},\;n} {\left[ {{{\bf{A}}_{{\bf{k}}n}}({\bf{r}}){{\hat a}_{{\bf{k}}n}}{e^{ -
i{\omega _{{\bf{k}}n}}t}} + {\bf{A}}_{{\bf{k}}n}^*({\bf{r}})\hat a_{{\bf{k}}n}^\dag {e^{i{\omega _{{\bf{k}}n}}t}}} \right]},
\end{split}
\end{equation}
where ${{\bf{A}}_{{\bf{k}}n}}({\bf{r}}) = {1 \mathord{\left/{\vphantom {1 {\sqrt {V{\omega _{{\bf{k}}n}}} }}} \right.
 \kern-\nulldelimiterspace} {\sqrt {V{\omega _{{\bf{k}}n}}} }} {{\bf{E}}_{{\bf{k}}n}}({\bf{r}})$ with $\textbf{E}_{\textbf{k}n}(\textbf{r})$
 being the Bloch eigenfunctions satisfying the following orthonormality condition:
\begin{equation}\label{eqn16}
\int\limits_V {d^3}r\varepsilon ({\bf{r}}){\bf{E}}_{{\bf{k}}n}({\bf{r}}){\bf{E}}_{{\bf{k}}'n'}^{*}({\bf{r}})  =
V{\delta_{{\bf{kk'}}}}{\delta_{nn'}},
\end{equation}
where $\textit{V}$ is the normalization volume of a PC. These eigenfunctions can be expanded as
\begin{equation}
{\bf{E}}_{{\bf{k}}n} ({\bf{r}}) = \sum_{\bf{G}}{{\bf{E}}_{{\bf{k}}n} ({\bf{G}})e^{i\left( {{\bf{k}} + {\bf{G}}} \right) \cdot {\bf{r}}} },\notag
\end{equation}
in which ${\bf{E}}_{{\bf{k}}n}({\bf{G}})$ are the Bloch eigenfunctions in the momentum representation.

\begin{figure}[]
\centering
\includegraphics[width=0.50\linewidth]{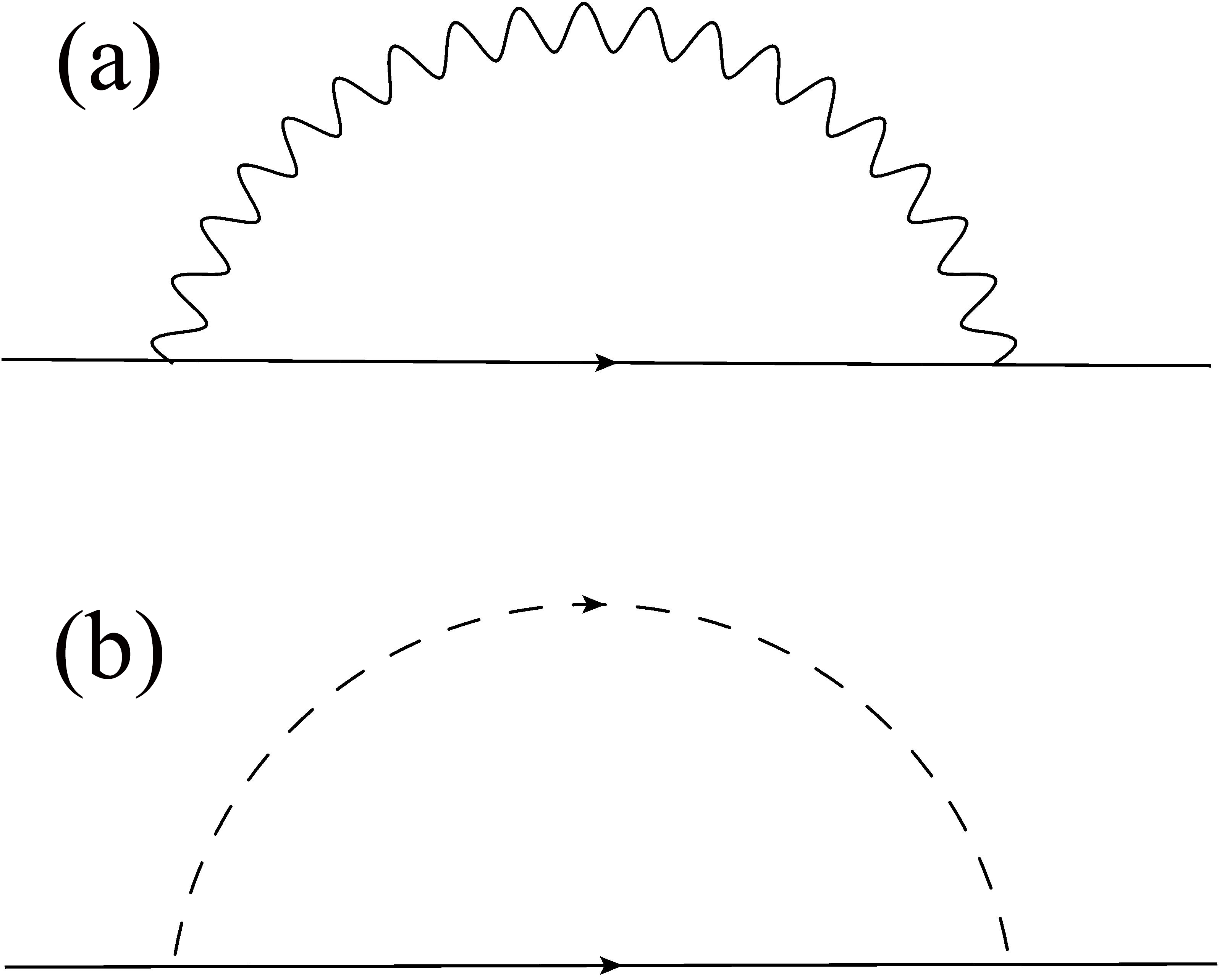}
\caption{\label{SelfEnDiagrammfig} The self-interaction diagram of an electron with (a) virtual photons in vacuum and (b) photon Bloch modes of
the PC medium}
\end{figure}

The PC correction to the electron mass, being a result of the modification of the interaction of an electron with its own electromagnetic
field in the PC medium, is a difference $\Delta E_{em}^{pc}(\textbf{p})$ between the self-energy corrections to kinetic energy in the PC
medium and in vacuum (Fig.~\ref{SelfEnDiagrammfig}) \cite{Gainutdinov2012}
\begin{equation}\label{eqn17}
\begin{split}
{{\Delta }{E_{em}^{pc}}}(\textbf{p}) = {\sum\limits_{{\bf{p}}'}\sum\limits_{{\bf{k}},\;{{n}}} {\frac{{\left\langle {\bf{p}}
\right|\widehat{H}_I^{pc}\left| {{\bf{p}}';{\bf{k}},n} \right\rangle \left\langle {{\bf{p}}';{\bf{k}},n} \right|\widehat{H}_I^{pc}\left| {\bf{p}} \right\rangle
}}{{\frac{{{{\bf{p}}^2}}}{{2{m_e}}} - \frac{{{\bf{p}}{'^2}}}{{2{m_e}}} - {\omega _{{\bf{k}}n}}}}} } \\
- {\sum\limits_{{\bf{p}}'}\sum\limits_{{\bf{k}},\;{\lambda }} \frac{{\left\langle {\bf{p}} \right|\widehat{H}_I^{}\left|
{{\bf{p}}';{\bf{k}},{\bm{\varepsilon}_\lambda }} \right\rangle \left\langle {{\bf{p}}';{\bf{k}},{\bm{\varepsilon}_\lambda }} \right|\widehat{H}_I^{}\left|
{\bf{p}} \right\rangle }}{{\frac{{{{\bf{p}}^2}}}{{2{m_e}}} - \frac{{{\bf{p}}{'^2}}}{{2{m_e}}} - \left| {\bf{k}} \right|}}}.
\end{split}
\end{equation}
If we take into account, that in the nonrelativistic limit, the Hamiltonian $\widehat{H}_I$ describing the interaction of the electron with the
electromagnetic field can be written in the form $\widehat{H}_I=-\frac{e}{m_e}{\widehat{\textbf{p}}} \cdot \widehat {\bf{A}}({\bf{r}})$, and the
PC-medium Hamiltonian is modified by replacing $\widehat {\bf{A}}({\bf{r}})$ with $\widehat {\bf{A}}_{pc}({\bf{r}})$, then we can rewrite Equation
(\ref{eqn17}) (see Section 1 in the Supplementary material for details \cite{SupplMater}) as
\begin{equation}\label{eqn18}
\begin{split}
{{\Delta }{E_{em}^{pc}}}(\widehat{\bf{{p}}}) = - \frac{\alpha }{{{2m_e^2\pi ^2}}}\left( \sum\limits_{\lambda } {\int\limits_{}
{\frac{{{d^3}{\bf{k}}}}{{2\left| {\bf{k}} \right|}}\frac{{{{\left| {\widehat{\bf{p}} \cdot {\bm{\varepsilon} _\lambda }({\bf{k}})}
\right|}^2}}}{{\frac{{{\widehat{\bf{p}}^2}}}{{2{m_e}}} - \frac{{{{\left( {\widehat{\bf{p}} - {\bf{k}}} \right)}^2}}}{{2{m_e}}} - \left| {\bf{k}}
\right|}}{\mkern 1mu} } } \right. - \\
- \left.{{\sum\limits_{n,\;{\bf{G}}} {\int\limits_{FBZ} {\frac{{{d^3}{\bf{k}}}}{{{\omega _{{\bf{k}}n}}}}\frac{{{{\left| {\widehat{\bf{p}}
\cdot {{\bf{E}}_{{\bf{k}}n}}\left( {\bf{G}} \right)} \right|}^2}}}{{\frac{{{\widehat{\bf{p}}^2}}}{{2{m_e}}} - \frac{{{{\left( {\widehat{\bf{p}} - {\bf{k}} -
{\bf{G}}} \right)}^2}}}{{2{m_e}}} - {\omega _{{\bf{k}}n}}}}{\mkern 1mu} } } }} \right) = \\
= - \frac{\widehat{\bf{p}}^2}{{{2m_e^2}}}\cdot (\Delta m_{em}^{pc}\left(\widehat{\bf{I}}_{\bf{p}} \right) - \Delta m_{em})
+ O{\left( \frac{{\bf{k}}^2}{2m_e} \right)} + O{\left( {\left| \widehat{\bf{p}}\right|} \right)^3}.
\end{split}
\end{equation}

It may be noted that the dependence of the matrix elements in Equation~(\ref{eqn18}) on the position of an electron in a PC disappeared. This
happened after integrating over $r$ in Equation~(3) (see the details in Section 1 of the Supplementary material \cite{SupplMater}). This fact makes the PC
correction of a free electron coordinate-independent. At the same time, when we calculate the Lamb shift, the dependence on the position may take a place
because, in this case, we deal with an atomic electron. Its wave function is now localized in the vicinity of a nucleus with coordinate
$r_0$ and is immersed in a Bloch electromagnetic wave described by a vector potential \cite{Scully1997quantum}. The corresponding matrix element
can be represented in the form
\begin{equation}
\begin{split}
\left\langle {i;{\bf{k}},n} \right|\widehat H_I^{pc}\left| j \right\rangle  =  - \frac{e}{m_e}\int {{d^3}r} \Psi _i^*({\bf{r}})
\left( {\widehat {\bf{p}} \cdot {\bf{A}}_{{\bf{k}},n}^*({\bf{r}})} \right)\Psi _j^{}({\bf{r}}) \approx  \\
\approx - \frac{e}{m_e}{{\bf{p}}_{ij}}
{\bf{A}}_{{\bf{k}},n}^*({{\bf{r}}_0}), \notag
\end{split}
\end{equation}
where $\left| i \right\rangle$ and $\left| j \right\rangle$ are atomic states. The dependence of the matrix elements on the coordinate $r_0$ is
a reason why the Lamb shift of atoms placed in the PC medium is position-dependent \cite{Wang2004giant,Sakoda2004}.
The first term on the right-hand part of Equation~(\ref{eqn18}) is just the ordinary low-energy part of the self-energy of an electron in vacuum,
which appears in the second-order perturbation theory \cite{Bjorken1965relativistic,Schweber2011}, whereas the second term is the modified
self-energy in the PC medium \cite{Gainutdinov2012}.

It is important that at the photon energies higher than a few tens of eV the refractive indices approach 1, and hence the contributions from
these terms cancel each other. In the PC medium, vacuum becomes anisotropic and $E^2 - \widehat{\bf{p}}^2$ is an invariant only under
the Lorentz boost with the velocity $\bf{v}$ directed along the direction ${\widehat{\bf{I}}_{\bf{{p}}}}$ of the electron momentum.
The "invariant" depends on this direction, namely
\begin{equation}\label{eqn5}
E^2-\widehat{{\bf{p}}}^2 = m_{pc}^2(\widehat {\bf{I}}_{\bf{p}}),
\end{equation}
where $m_{pc}(\widehat {\bf{I}}_{\bf{p}}) = m_0 + m_{em} + \delta m_{pc}(\widehat {\bf{I}}_{\bf{p}})$. The electron mass correction $\Delta m_{em}$
defined in Equation~(\ref{eqn4}) is finite only because of the cutoff $k_0$, and there is no renormalization procedure that can be used to remove
such cutoffs. This means that the ultraviolet divergences in equations describing the electromagnetic mass of the electron are not renormalizable,
and the only way to solve the problem is to include the electromagnetic mass into the observable mass $m_e$. In the case where we deal with an
electron interacting with the PC vacuum, only the part of the electromagnetic mass that is equal to the electromagnetic mass $\Delta m_{em}$ can be
hidden in the observable electron mass $m_e$. Only the correction $\delta m_{pc}(\widehat {\bf{I}}_{\bf{p}})$ remains unhidden. In this way we arrive
at the operator of the correction to the electron mass (see Equation~(\ref{eqn18}))
\begin{equation}\label{eqn6}
\begin{split}
{\delta {m}}_{pc}\left(\widehat{\bf{I}}_{\bf{p}} \right)  = \Delta m_{em}^{pc}\left(\widehat{\bf{I}}_{\bf{p}} \right) - \Delta m_{em} = \\
= \frac{\alpha }{{\pi ^2 }}\left[ {\sum\limits_{n}  \int\limits_{FBZ}  \,\frac{{d^3 {\bf{k}} }}{{\omega
_{{\bf{k}}n}^2 }}\sum\limits_{\bf{G}} {\left| {{\widehat {\bf{I}}_{\bf{p}}}\cdot {\bf{E}}_{{\bf{k}}n} ({\bf{G}})} \right|^2 } } \right. -  \\
- \left.{\int {\frac{{d^3 {\bf{k}}}}{{2{\bf{k}}^2 }}}\sum\limits_{\lambda  = 1}^2  \,\mathop {\left| { {\widehat {\bf{I}}_{\bf{p}}}\cdot
\bm{\varepsilon} _\lambda  ({\bf{k}})} \right|}\nolimits^2 } \right]. \\
\end{split}
\end{equation}
Therefore, the operator of the correction ${\delta m}_{pc}\left({\widehat{\bf{I}}}_{\bf{p}} \right)$ to the electron mass is the result of the subtraction
of the electromagnetic mass $\Delta m_{em}$ in vacuum from the electromagnetic mass $\Delta m_{em}^{pc}\left(\widehat{\bf{I}}_{\bf{p}} \right)$
in the PC medium. As a consequence, the PC correction to the electron mass is free from ultraviolet divergences (for details, see Section 3 in the
Supplementary material \cite{SupplMater}). A similar regularization method has been recently studied in the problem of free-electron radiation
into a medium, known as Cherenkov radiation \cite{Roques2018}.

It follows from Equation~(\ref{eqn6}), that the PC correction ${\delta {m}}_{pc}\left(\widehat{\bf{I}}_{\bf{p}} \right)$ is independent of the
atomic potential or the periodic potentials in solids and depends only on the interaction of an electron with its own radiation field.
The interaction of an electron atom with its own radiation field consists of both the interaction of each atomic electron with its own
radiation field and the self-interaction of the atom involving the Coulomb interaction of the atomic electrons with the nucleus. In the case of an
atom in free space, the first interaction process generates the electromagnetic mass of the electron, whereas the second one gives rise to the Lamb
shifts of the atomic energy levels. The difference of the energy levels of atoms is constant in the case of vacuum and should be shifted in any
isotropic media. When we deal with an atom placed in the PC medium, its energy levels have a shift dependent on the atomic electron state. The energy
level shift related to the self-energy interaction of an atomic electron is much larger than the Lamb shift, i.e., the correction to
the Coulomb interaction of the atomic electron with the nucleus.

The Hamiltonians of atoms in the PC medium must be completed with the operators ${\delta {m}}_{pc}\left(\widehat{\bf{I}}_{\bf{p}} \right)$ for each
electron. For example, the Hamiltonian of the atomic hydrogen takes the form
\begin{equation}\label{eqn10}
\widehat{H}_{pc} = {\delta {m}}_{pc}\left(\widehat{\bf{I}}_{\bf{p}} \right) + \widehat{H},
\end{equation}
where $\widehat{H}$ is the Hamiltonian of the atomic hydrogen in free space containing the rest mass part, which is not usually considered.
The atomic states and energies are thus determined by the equation
\begin{equation}\label{eqn11}
\widehat{H}_{pc}\left| {{\Psi _{i,pc}}} \right\rangle  = {E_{i,pc}}\left| {{\Psi _{i,pc}}} \right\rangle.
\end{equation}
It can be solved perturbatively expanding the solution in powers of ${\delta {m}}_{pc}\left(\widehat{\bf{I}}_{\bf{p}} \right)$. At
leading order we get $\left| {{\Psi _{i,pc}^{(1)}}}
\right\rangle  = \left| {{\Psi _{i}}} \right\rangle$ and $E_i^{(1)} = \left\langle {{\Psi _i}}\right|{\delta {m}}_{pc}\left(\widehat{\bf{I}}_{\bf{p}} \right)
\left| {{\Psi _i}} \right\rangle  + {E_i}$, where ${E_i}$ is the energy of the state $\left| {{\Psi _i}}\right\rangle$ of the atom in the free space.
It should be noted that the correction $E^{(1)}_i$ - $E_i$ depends only on the orbital $l$ and the magnetic $m_l$
quantum numbers, but not on the electron coordinate: $\left\langle {{\Psi}} \right|{\delta m}_{pc}\left(\widehat{\bf{I}}_{\bf{p}} \right)\left| {{\Psi}}
\right\rangle = \left\langle {{l,m_l}}\right|{\delta m}_{pc}\left(\widehat{\bf{I}}_{\bf{p}} \right)\left| {{l,m_l}}
\right\rangle$. For the atoms of the hydrogen and alkali metals $l$ = 0, $m_l = 0$. Hamiltonian of an $\textit{N}$-electron atom placed in the PC
medium is described in a similar manner:
\begin{equation}\label{eqn12}
\widehat{H}_{pc}^N = \sum\limits_{i = 1}^N {{\delta m}_{pc}^{(i)}}\left(\widehat{\bf{I}}_{\bf{p}} \right)  + \widehat{H}_{}^N,
\end{equation}
where $\widehat{H}^N$ also contains the rest masses of each electron. Since these parts are always constant they do not appear in the energy of the
transition between states of the atom in vacuum. In a PC, however, there are corrections to these energies that could significantly modify
familiar optical spectra of atoms. Transitions between states of the $N$-electron atom are accompanied by changes of the configurations of
electrons in the subshells, and the PC-medium corrections to their energies could have complex structure.

\section{Ionization energy of atoms in one-dimensional photonic crystals}\label{S3}

\begin{figure}[]
\centering
\includegraphics[width=1\linewidth]{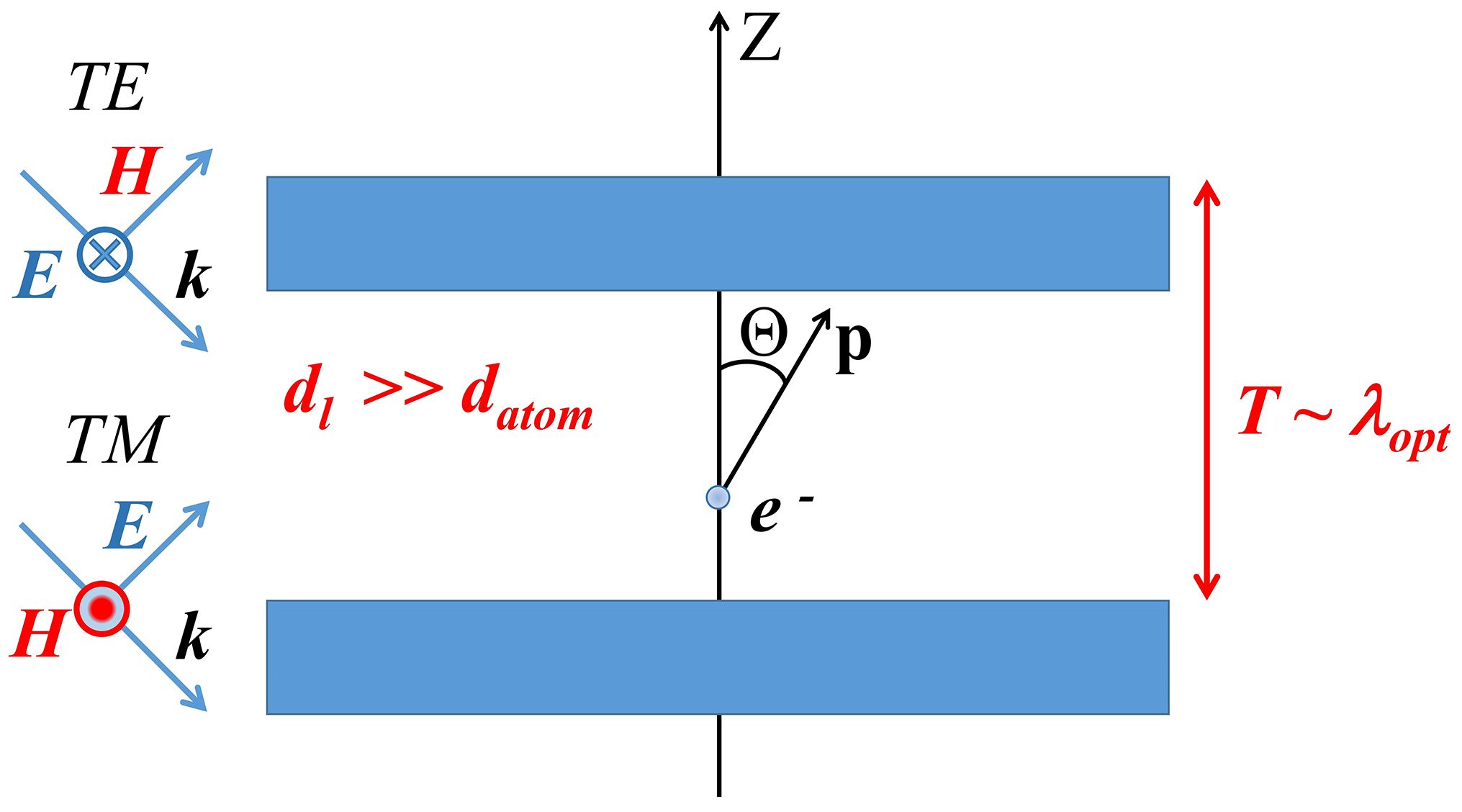}
\caption{\label{1DPCmodelfig} A free electron placed in an air void of the one-dimensional PC having cylindrical symmetry. The electron
propagates at an angle $\Theta$ to the $\textit{Z}$\;-\;axis of the crystal. The size of the vacuum layer of the PC is $10^3$ times larger than
the average size of the atom. In the simplest case, we can put a PC in a sealed flask and measure the reaction rate of gas-phase reactants pumped
into the voids of a large enough high-quality PC.}
\end{figure}

Let us consider one-dimensional PCs (Fig.~\ref{1DPCmodelfig}) because they are most important for applications of the effect under study.
In the case of a one-dimensional PC with a given $\textit{Z}$\;-\;axis, the coefficients ${\bf{E}}_{{\bf{k}}n} ({\bf{G}})$ have the
polarization structure
\begin{equation}\label{eqn8}
{{\bf{E}}_{{\bf{k}}n}}(\textbf{G}) = \sum\limits_{\lambda = 1}^2 {{E}_{{\bf{k}}n\lambda}({G})}{{\bm{\varepsilon }}_{\lambda}
(\bf{k_G})},
\end{equation}
where ${\bm{\varepsilon}}_{{1}}(\bf{k_G}) $ and ${\bm{\varepsilon}}_{{2}}(\bf{k_G}) $ are unit vectors of the TE (transverse electric) and TM
(transverse magnetic) polarization, respectively, ${\bf{k_G}} = {\bf{k}} + G{{\bf{e}}_z}$.

\noindent Thus, for an one-dimensional PC the operator of the self-energy correction (\ref{eqn6}) can be rewritten as (see the details in
Section 2 of the Supplementary material \cite{SupplMater})
\begin{equation}\label{eqn9}
{\delta {m}}_{pc}\left(\widehat{\bf{I}}_{\bf{p}} \right) = A + \left( {{{\widehat {\bf{I}}}_{\bf{p}}} \cdot {{\widehat {\bf{I}}}_{pc}}} \right)^{2}B,
\end{equation}
where ${{{\widehat {\bf{I}}}_{pc}}}$ is the unit vector of the 1D PC axis, which coincides with the vector ${\bf{e}}_z$, and
\begin{equation}
\begin{split}
A = \frac{{{\alpha }}}{{{\pi }}}\sum\limits_{n,\;G}^{} {\int {{k_\rho }d{k_\rho }} \int\limits_{FBZ}^{} {d{k_z}\left( {\frac{{{{\left|
{E_{{\bf{k}}n1}^{}\left( G \right)} \right|}^2}}}{{{{\omega }}_{{\bf{k}}n1}^2}}\frac{{k_{Gz}^2}}{{k_\rho ^2 +k_{Gz}^2}}}\right.}} + \\
 + {{\left.\frac{{{{\left| {E_{{\bf{k}}n2}^{}\left( G \right)} \right|}^2}}}{{{{\omega }}_{{\bf{k}}n2}^2}} \right)}}-\frac{{4{{\alpha
}}}}{{3{{\pi }}}}{\int {dk}}\notag,
\end{split}
\end{equation}
\begin{equation}
\begin{split}
B = \frac{{{\alpha }}}{{{\pi }}}\sum\limits_{n,\;G}^{} {\int {{k_\rho }d{k_\rho }} \int\limits_{FBZ}^{} {d{k_z}\left( {\frac{{{{\left|
{E_{{\bf{k}}n1}^{}\left( G \right)} \right|}^2}}}{{{{\omega }}_{{\bf{k}}n1}^2}}\frac{{2k_\rho ^2 - k_{Gz}^2}}{{k_\rho ^2 + k_{Gz}^2}} }\right.}}- \\
- {{\left.{\frac{{{{\left| {E_{{\bf{k}}n2}^{}\left( G \right)} \right|}^2}}}{{{{\omega }}_{{\bf{k}}n2}^2}}} \right)}}\notag. \\
\end{split}
\end{equation}
Here, $\omega _{{\bf{k}}n1}$ and $\omega _{{\bf{k}}n2}$ are dispersion relations for the TE and TM Bloch modes satisfying transcendental
equation \cite{Skorobogatiy2009}.

\begin{equation}\label{eqn19}
\begin{split}
 \cos \left( {k_z \left( {d_h  + d_l } \right)} \right) = \cos \left( {k_z^h d_h } \right)\cos \left( {k_z^l d_l } \right)  - \\
 - {{\left( {r_{ 1, \;2 }  + r_{ 1, \;2 }^{ - 1} } \right)} \mathord{\left/  {\vphantom {{\left( {r_{ 1, \;2 }  + r_{ 1, \;2 }^{ - 1} } \right)} 2}} \right.
 } 2} \cdot \sin \left( {k_z^h d_h } \right)\sin \left( {k_z^l d_l } \right), \\
 \end{split}
 \end{equation}
where ${r_1} = {{k_z^l} \mathord{\left/{\vphantom {{k_z^l} {k_z^h}}} \right.\kern-\nulldelimiterspace} {k_z^h}}, {r_2} = n_h^2{r_1}, k_z^i  =
\sqrt {{{{\omega }}_{{\bf{k}}n}^2} n _i^2  - k_\rho^2 }$
with $d_h$ and $d_l$ being the thicknesses of the layers of the one-dimensional PC with high (\textit{h}) and
low (\textit{l}) refractive indices $n_h$ and $n_l = 1$ (air voids), $k_\rho=\sqrt {k_x^2+k_y^2}$.
\begin{figure}[]\center
\begin{tabular}{cc}
\includegraphics[width=0.48\linewidth]{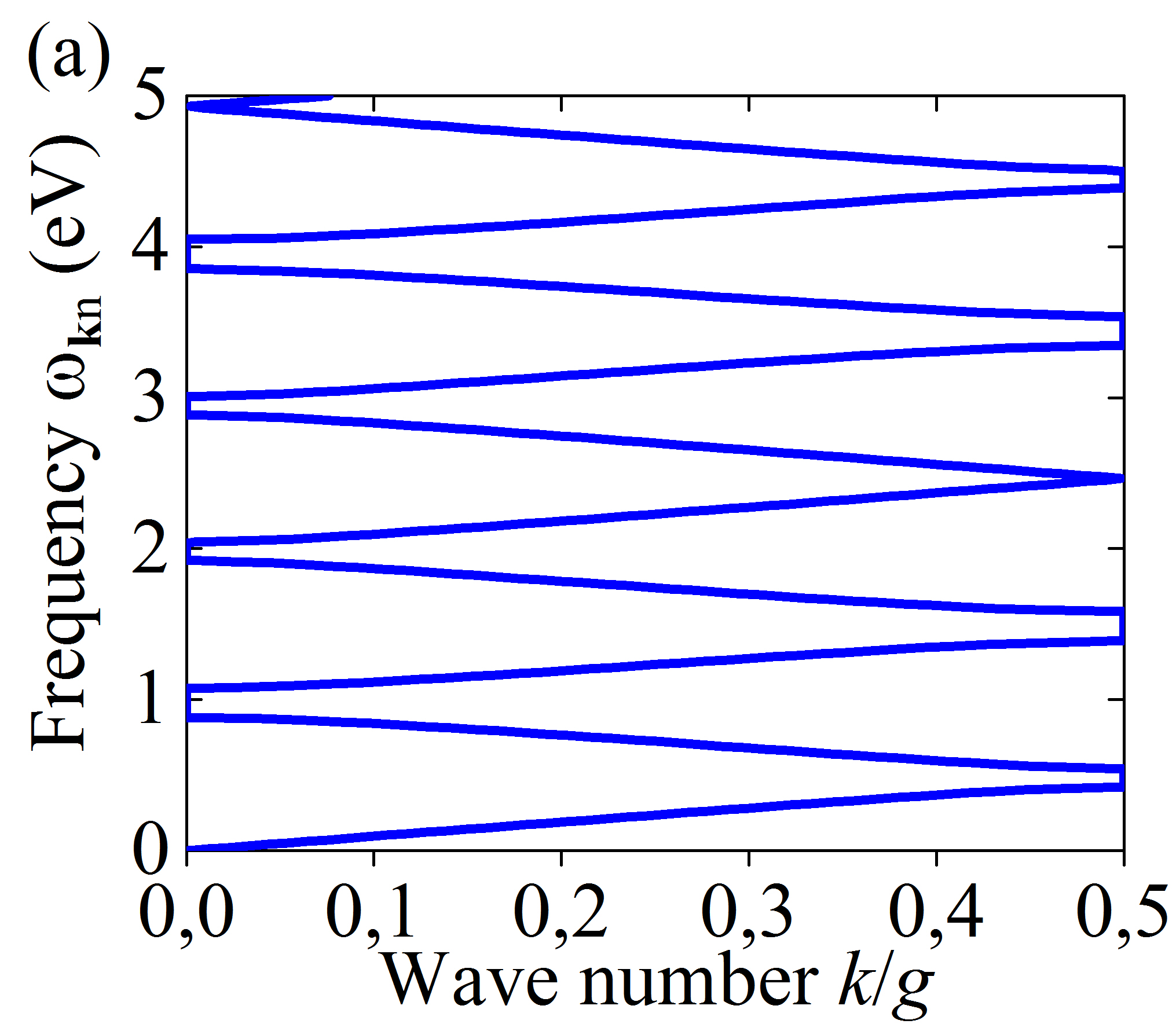}
&
\includegraphics[width=0.48\linewidth]{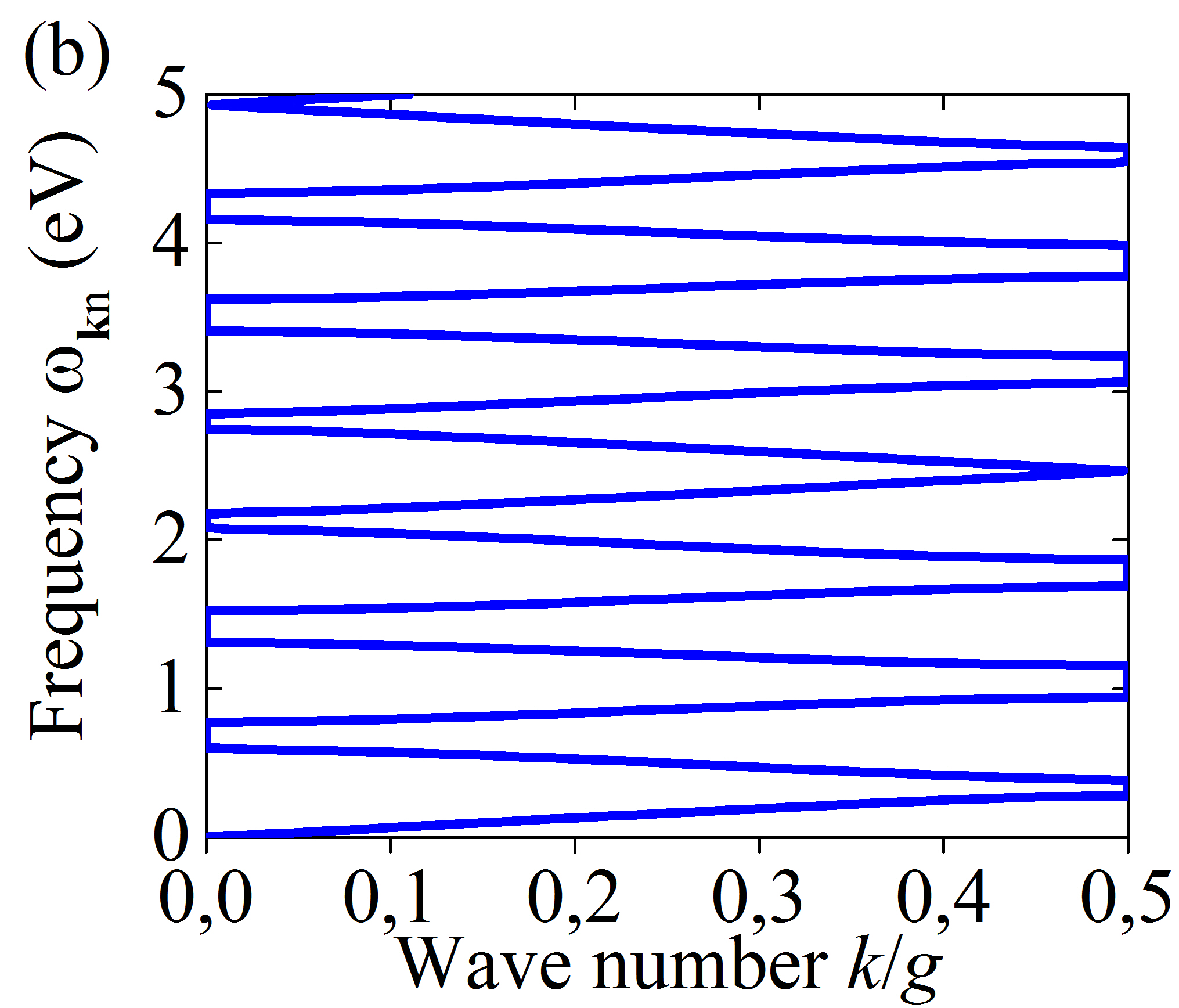} \\

\includegraphics[width=0.48\linewidth]{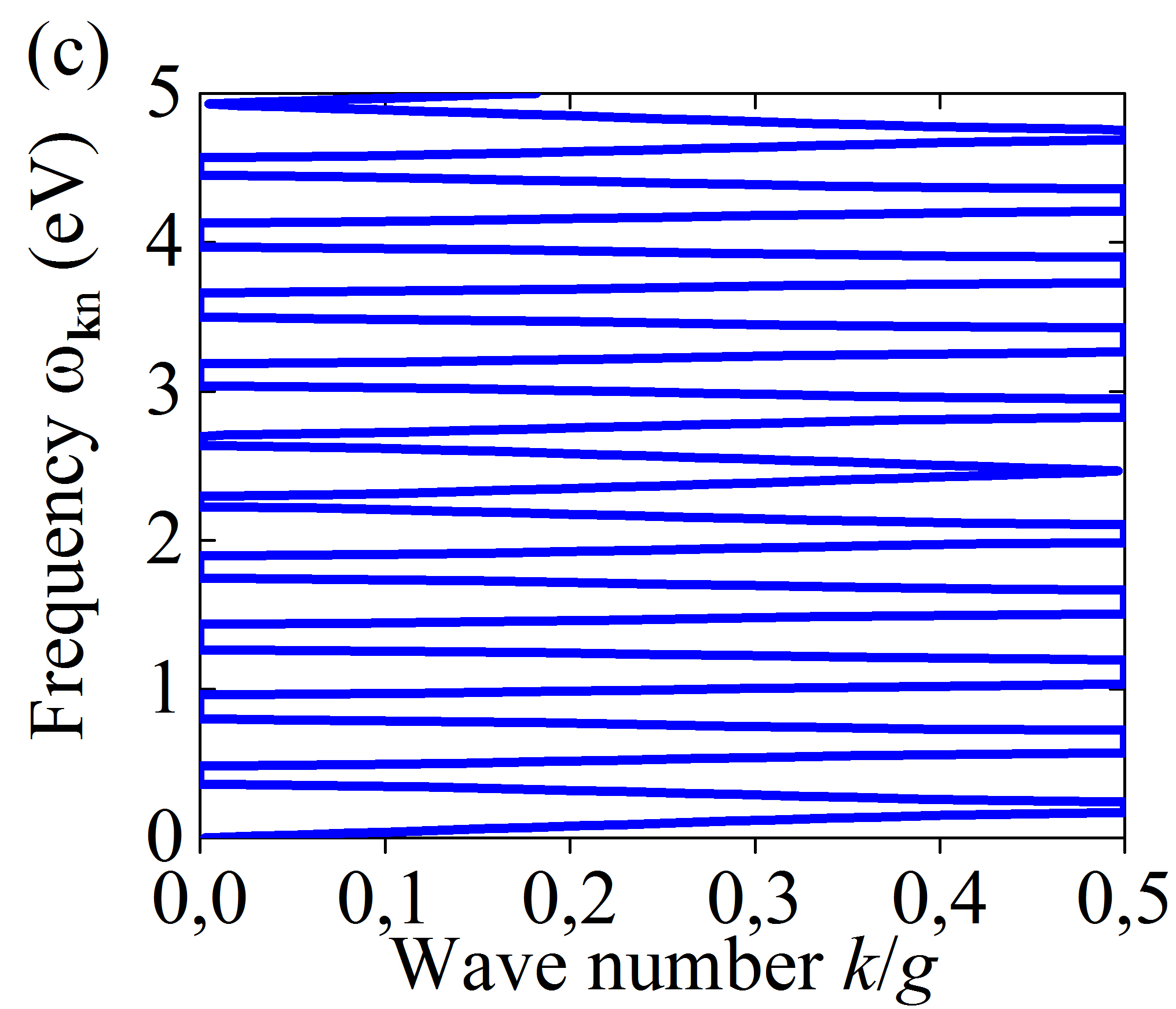}
&
\includegraphics[width=0.48\linewidth]{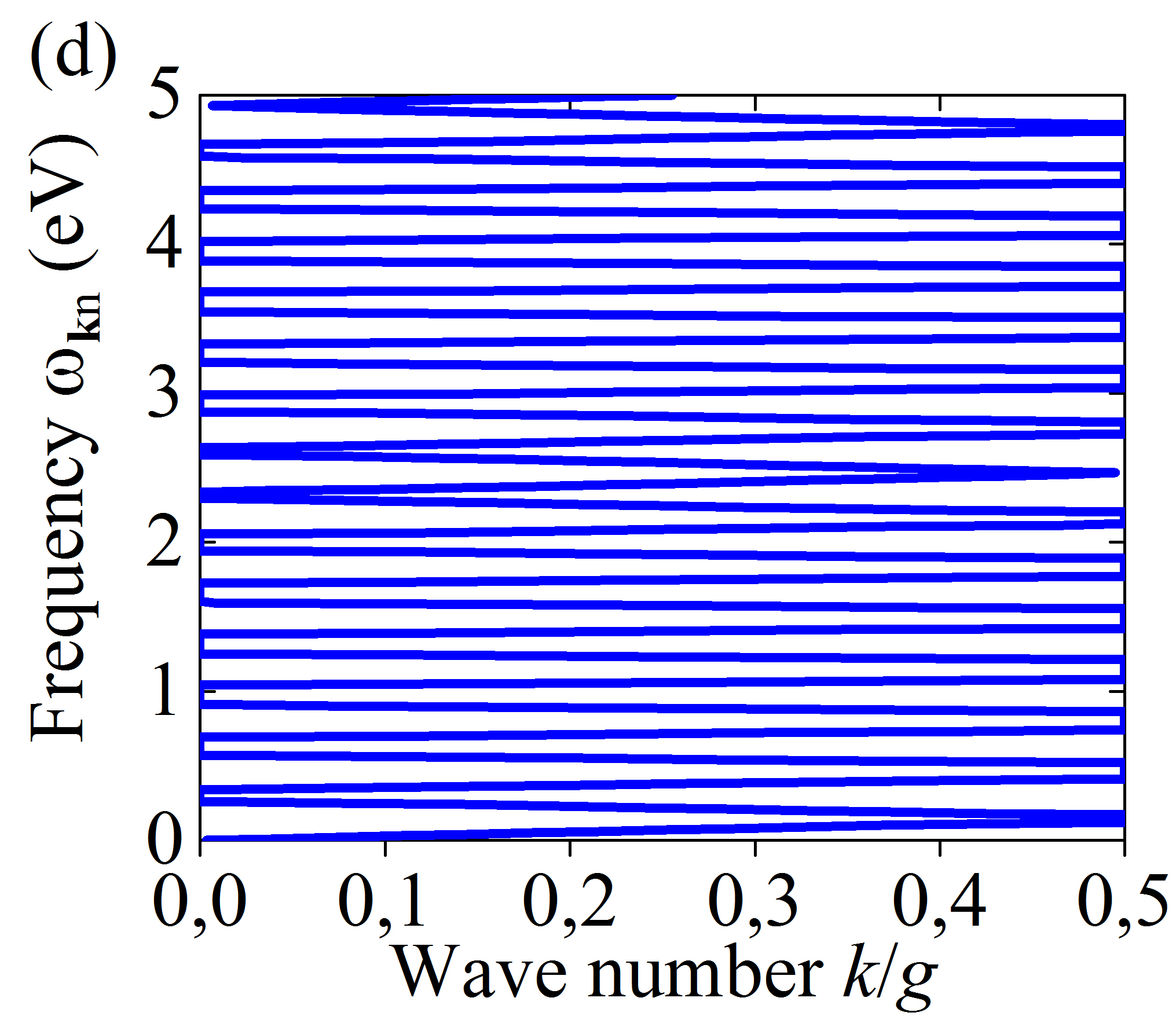}
\end{tabular}
\caption{\label{DispRelfig} Dispersion relations for TM Bloch modes in one-dimensional PC for different values
of the refractive index (a) $n_h$ = 2, (b) $n_h$ = 3, (c) $n_h$ = 5, (d) $n_h$ = 7.
The number of bands $n$ in dispersion relations linearly increases with the refractive index of PC material host.}
\end{figure}

Let us consider the ionization process for the atoms of hydrogen atom and alkali metals, where the transition is defined by a single valence electron.
In this case, the lower state is the ground state of an atom ($S$-state) and the upper state is the free state. The difference between energies of
these states determines the binding energy $E_{bind}$ of the electron. It should be noted that, in the case of the PC medium, $E_{bind}^{pc}$ depends on
the direction of the free-electron momentum $\bf{p}$, which can impact on the configurations of bonds in molecules. However, primarily, the PC medium corrections
modify the ionization energy of atoms, defined as the minimum energy necessary to remove an electron. According to this definition, the
correction to the ionization energy of an atom is determined by the equation
\begin{equation}\label{eqn13}
\delta E_{ion}^{pc} = \delta m_{pc}^{min} - {\delta m_{pc}^{{l,m}}},
\end{equation}
where $\delta m_{pc}^{min}$ is the smallest correction determining by Equation~(\ref{eqn9}), and
\begin{equation}
{\delta m_{pc}^{{l,m}}} = \left\langle {{\Psi}} \right|{\delta m}_{pc}\left(\widehat{\bf{I}}_{\bf{p}} \right)\left| {{\Psi}} \right\rangle =
\left\langle {{l,m_l}}\right|{\delta m}_{pc}\left(\widehat{\bf{I}}_{\bf{p}} \right)\left| {{l,m_l}} \right\rangle. \notag
\end{equation}
For the atoms of hydrogen and alkali metals, we have $l$ = 0, $m_l$ = 0.

\begin{figure}[]
\centering
\includegraphics[width=1\linewidth]{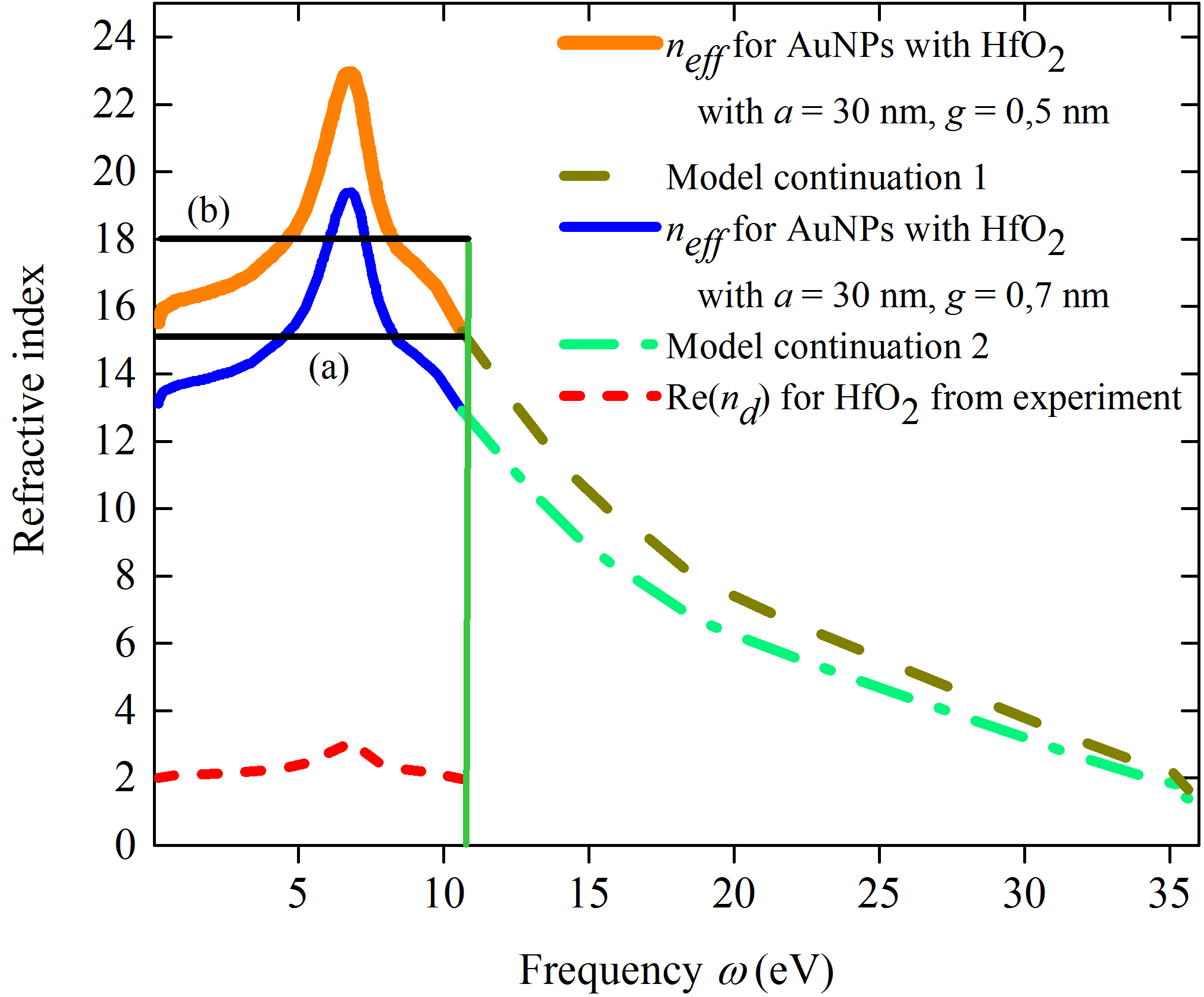}
\caption{\label{nhavermetafig} The spectral dependencies $n_{eff}(\omega)$ for a metamaterial consisting of an AuNPs ensemble coated with HfO${}_{2}$ (thick
orange and blue solid lines) extracted from the experiment (${n_d}\left( \omega  \right) \approx \sqrt {{\varepsilon _d}\left( \omega
\right)} $ of the HfO${}_{2}$) (red dashed line) \cite{Franta2015}. The curves for the same region of frequencies as the experimental one have been plotted using equation ${n_{eff}}\left( \omega  \right) = {[(a/g){\varepsilon _d}\left( \omega  \right)]^{1/2}}$
(see \cite{Chung2016}), with $\textit{a}$ = 30 nm, $g$ = 0.5 nm (thick orange solid line), and $\textit{a}$ = 30 nm, $g$ = 0.7 nm (blue solid
line). The rest parts of the curves have been chosen to provide the fact that at the high frequencies $n_{eff}$$\rightarrow$1
(brown dashed line and green dash-dotted line). The average refractive indexes (a) $n_{eff}$ = 15 and (b) $n_{eff}$ = 18 are given for the
first and second spectral line at maximum frequency $\omega _{{\bf{k}}n}^{\max }$ = 10.65 eV.}
\end{figure}

\begin{figure}[]
\centering
\includegraphics[width=1\linewidth]{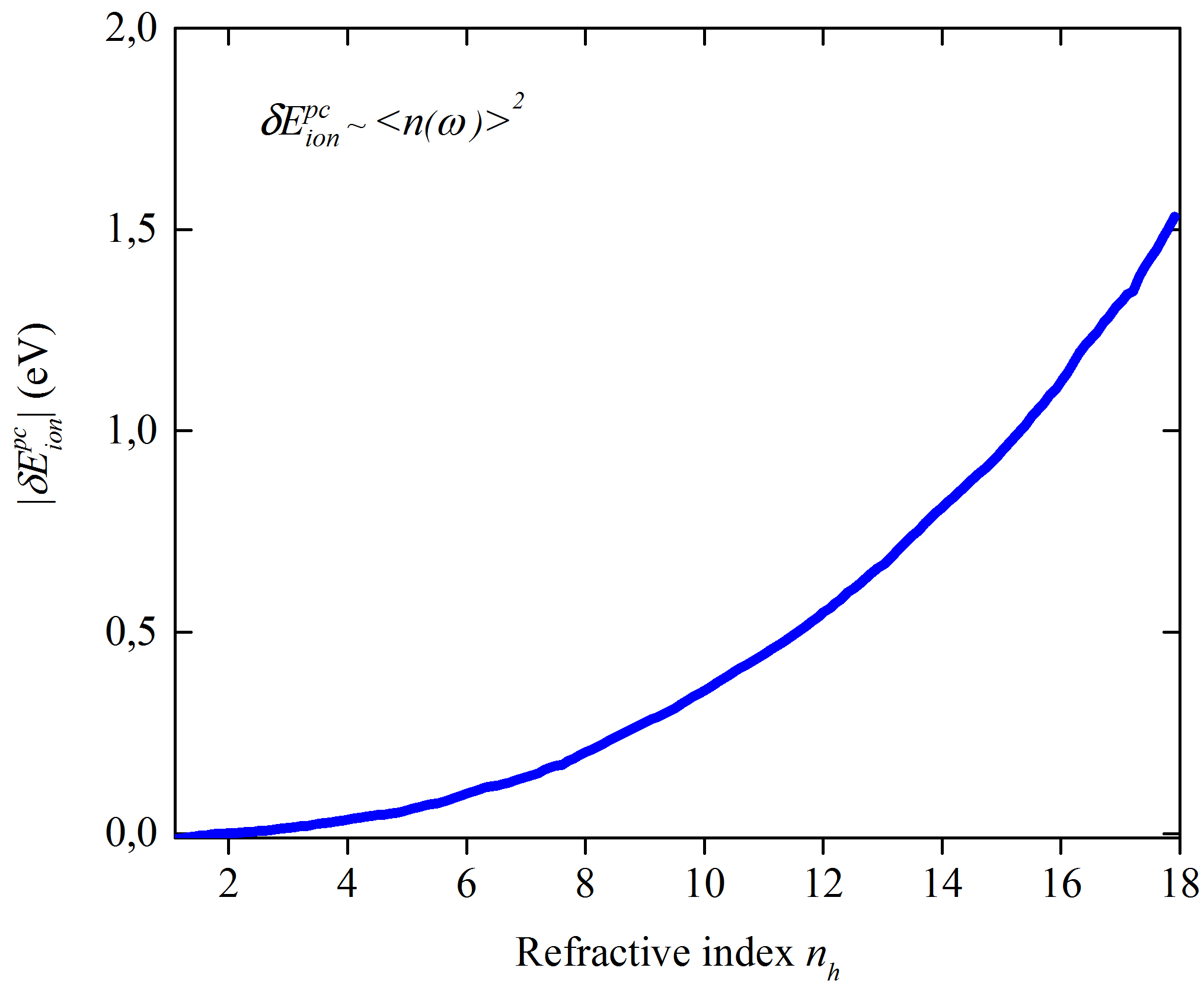}
\caption{\label{FullIntfig} The dependence of the absolute value of $\delta E_{ion}^{pc}$ on the averaged refractive index $n_h$ of high-index layers of
one-dimensional PC. We use $\omega _{{\bf{k}}n}^{\max }$ = 10.65 eV in dispersion relations. This dependence does not account for the material dispersion.
At the same time, we take into account the material dispersion in the calculation of the ionization energy correction.}
\end{figure}

After substituting Equations~(\ref{eqn6}) and (\ref{eqn9}) into Equation~(\ref{eqn13}), the ionization energy correction can be represented in the form:
\begin{equation}\label{eqn14}
\begin{split}
\delta E_{ion}^{pc} = -\frac{2\alpha }{{3\pi }}\sum\limits_{n,\;G}{\left[ {\int\limits {{k_\rho }d{k_\rho }} \int\limits_{FBZ} {d{k_z}} }
\right.} \left( {\frac{{{{\left| {E_{{\bf{k}}n1} \left( G \right)} \right|}^2}}}{{\omega _{{\bf{k}}n {1} }^2}}}\cdot \right. \\
\left.\cdot {\left. {\frac{{k_{Gz}^2 - 2k_\rho ^2}}{{k_\rho ^2 + k_{Gz}^2}} + \frac{{{{\left| {E_{{\bf{k}}n2} \left( G \right)}
\right|}^2}}}{{\omega _{{\bf{k}}n{2} }^2}}} \right)} \right].
\end{split}
\end{equation}
When $\textbf{k} \to \infty$, the PC medium can be considered as an isotropic free space with dielectric constant $\varepsilon(\textbf{r}) = 1$
\cite{Wang2004giant}. From this it follows, that the integrand of Equation~(\ref{eqn14}) tends to zero.
The absolute value of $\delta E_{ion}^{pc}$ increases significantly with the refractive index of the PC material host (Fig.~\ref{nhavermetafig}).
We can easily verify the dependence for $\delta E_{ion}^{pc}$ (Fig.~\ref{FullIntfig}) by simplifying Equation~(\ref{eqn14}) and averaging the
refractive index $n_h(\omega)$. This is due to the fact that the number of bands $n$ in dispersion relations linearly increases with the refractive index
of PC material host (Fig.~\ref{DispRelfig}).
An additional dependence of the correction $\delta E_{ion}^{pc}$ on the refractive index arises because of the subsequent integration by $k_{\rho}$.
It is important that increasing the number of allowed states (the photonic density of states)  with the refractive index leads to significant
modification of the self-energy interaction of an electron with its own radiation field.

Thus, the materials with a highly tunable refractive index in a wide spectral range of light are needed.
Appropriate refractive properties are demonstrated by optical thin films formed by nonabsorbing dielectric material such as HfO${}_{2}$ (refractive
index $n_d$ $\sim$ 2 to 3 up to 10 eV) \cite{Franta2015}, and we consider only real part of refractive index of the thin films HfO${}_{2}$.
At the same time, it has been recently shown \cite{Lee2015,Chung2016,Kim2016,Kim2018} that the precise control
over the geometrical parameters of a nanoparticle superlattice monolayer leads to a dramatic increase in refractive index, far beyond the naturally
accessible regime. According to the theory of the optical effective media \cite{Chung2016}, which is in good agreement with the experiment
\cite{Kim2016,Kim2018}, the effective refractive index of such metamaterial is determined by the formula $n_{eff} = [(a/g)\varepsilon_d]^{1/2}$,
with array period $a$, the gap between particles $g$, and permittivity of the gap-filling dielectric $\varepsilon_d$. We assume that metamaterial
consisting of Au nanoparticles (AuNPs) ensemble coated with HfO${}_{2}$ having $a$ = 30 nm, $g$ = 0.7 nm and $a$ = 30 nm, $g$ = 0.5 nm allows one to achieve the unnaturally highly tunable refractive index (Fig.~\ref{nhavermetafig}).

\begin{figure}
\centering
\includegraphics[width=1\linewidth]{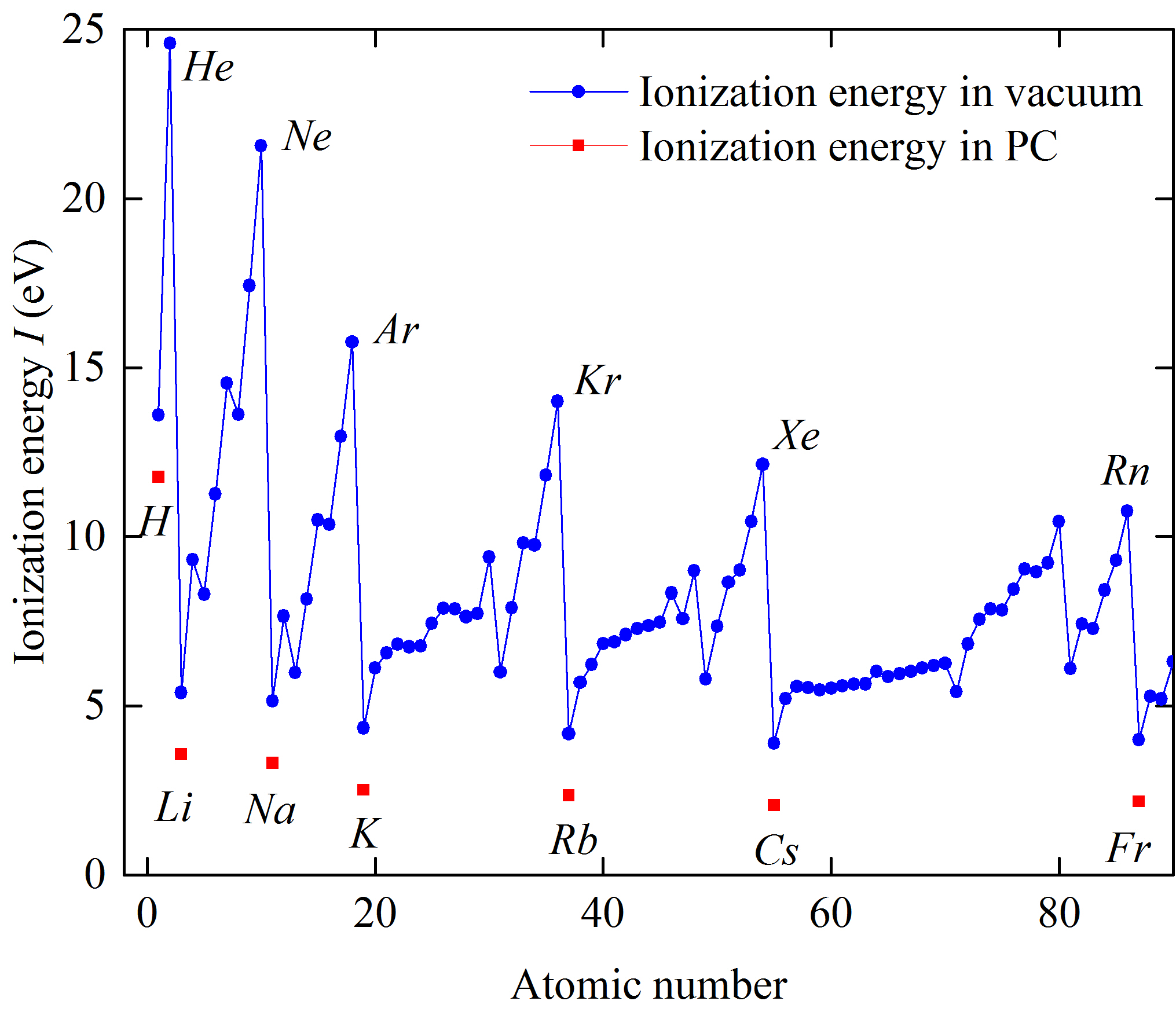}
\caption{\label{PeriodTable182fig} Comparison of the ionization energy of the atoms hydrogen and alkali metals in the case of vacuum (blue dots)
 and the PC medium (red squares) based on a metamaterial with $\textit{a}$ = 30 nm and $g$ = 0.7 nm. The ionization energy correction
 $\delta E_{ion1}^{pc}$ = $-$ 1.82 eV.}
\end{figure}

\begin{figure}
\centering
\includegraphics[width=1\linewidth]{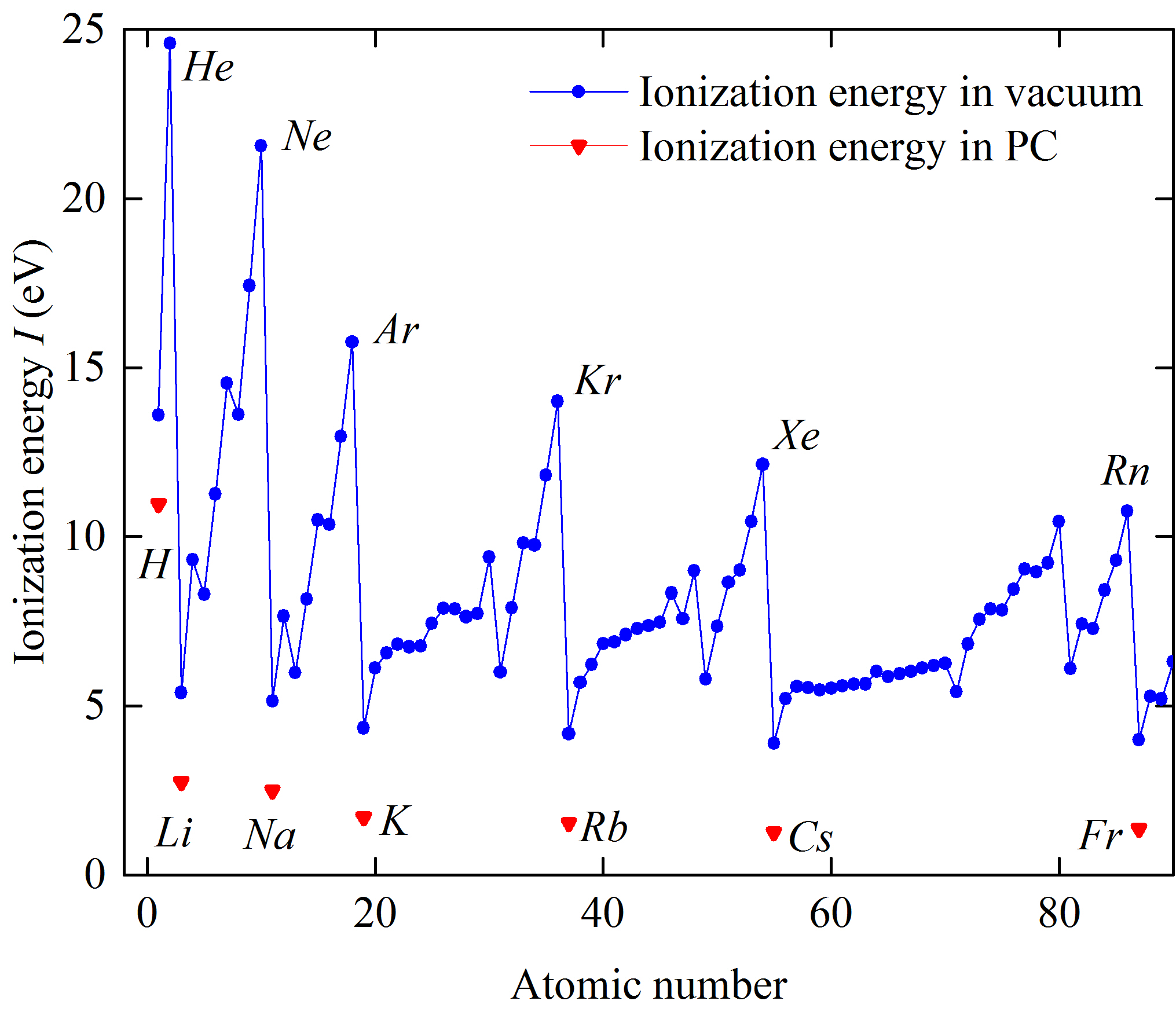}
\caption{\label{PeriodTable264fig} Comparison of the ionization energy of the atoms hydrogen and alkali metals in the case of vacuum (blue dots)
and the PC medium (red triangles) based on a metamaterial with $\textit{a}$ = 30 nm and $g$ = 0.5 nm. The ionization energy correction
$\delta E_{ion2}^{pc}$ = $-$ 2.64 eV.}
\end{figure}

To demonstrate the controllability of the ionization energies of the atoms hydrogen and alkali metals, we have estimated the ionization energy
corrections $\delta E_{ion}^{pc}$ of such atoms placed in voids of PC based on a high-index metamaterial with $a$ = 30 nm, $g$ = 0.7 nm, and $a$ = 30 nm, $g$ = 0.5 nm.
We have obtained the values $-$ 1.82 eV and $-$ 2.64 eV, respectively. The comparisons of the ionization energies in the case of vacuum and
the PC medium are represented in Fig.~\ref{PeriodTable182fig} and Fig.~\ref{PeriodTable264fig}. From these figures and
Table \ref{TableI}, we can see that the ionization energy correction in the PC medium could be comparable with the ionization energies of
atoms in free space.

As for the future experiment, our proposal is to put a PC in a sealed flask and measure the reaction rate of gas-phase reactants pumped into the voids of a large enough high-quality PC. Chemical reactions proceeding from reactants to a product
can be expressed as a transformation from a reactant energy minimum to a product energy minimum through a transition state. The temperature dependence of a chemical reaction provides important information on the reaction rate and energy barrier or activation energy \cite{Koga1994review,Smith2008temperature}.
Reaction rate constants for most chemical reactions closely follow an Arrhenius equation \cite{Ebbing2016general} of the form
\begin{equation}
k = Ae^{-E_a/RT},\notag
\end{equation}
where $A$ is the frequency factor, $E_a$ is activation energy, $R$ is the gas constant, $T$ is the absolute temperature. The activation energy and consequently reaction
rate of a particular chemical reaction are standard under ordinary conditions. Decreasing of ionization energies of reacting atoms placed in air voids of the PC medium leads to the significant decreasing of activation energy and acceleration of chemical reactions.
This effect is especially large when the PC consists of metamaterials with a highly tunable refractive index and voids.

\begin{table}[h]
\caption{\label{TableI} Ionization energy of hydrogen atom and the alkali metals in vacuum  \cite{NIST} and 1D PC}
\centering
\begin{ruledtabular}
\begin{tabular}{cccc}

   {\parbox[c]{1.5cm}{Chemical element}} & {\parbox[c]{3cm}{Standard ionization energy in vacuum \textit{I} ($eV$)}}  & \multicolumn{2}{c} {\parbox[c]{3.6cm}{Ionization energy in PC ($eV$)}} \\ \cline{3-4}
                                                        &                                                                                                      &{\parbox[c]{1.8cm}{$\delta E_{ion1}^{pc}$ = $-$ 1.82 $eV$}} & {\parbox[c]{1.8cm}{$\delta E_{ion1}^{pc}$ = $-$ 2.64 $eV$}}    \\
                                                        &                                                                                                       &                                                                            &                                                                  \\    \hline
    $H$  &13.60 & 11.78 & 10.96 \\
    $Li$ & 5.39 & 3.57 & 2.75 \\
    $Na$ & 5.14 & 3.32 & 2.50 \\
    $K$  & 4.34 & 2.52 & 1.70 \\
    $Rb$ & 4.18 & 2.36 & 1.54 \\
    $Cs$ & 3.90 & 2.08 & 1.26 \\
    $Fr$ & 4.07 & 2.25 & 1.43 \\
\end{tabular}
\end{ruledtabular}
\end{table}

\section{Conclusion}\label{S4}

In conclusion, we have shown that the modification of the interaction of an atom placed into the air voids of a PC with its own electromagnetic
field gives rise to the significant change in its ionization energy. The absolute value of the ionization energy correction increases
significantly with the refractive index of the material host of the PC medium. The effect is strongly enhanced when the PC is made from the
high-index metamaterials. Moreover, the effect is controllable. Controlling the geometrical parameters of a nanoparticle superlattice of metamaterial
can allow one to control chemical reactions that strongly depend on the atomic ionization energies. The effect under study allows one to come beyond the limitations put on by the
Periodic Table on physical and chemical processes and can open up new horizons for the synthesis of novel chemical compounds that could be used in
pharmaceutical and other medical-related activities.


\section*{Declaration of competing interest}

The authors declare that they have no known competing financial interests or personal relationships that could have appeared to influence the
work reported in this paper.

\section*{Acknowledgements}
The authors are deeply grateful for the useful discussions they have had with Dr. A. A. Akhmadeev (Kazan Federal University, Tatarstan
Academy of Sciences).

\newpage
\begin{center}
\large\textbf{Quantum electrodynamics in photonic crystals and controllability of ionization energy of atoms \\
(Supplementary material)}
\end{center}
\vspace{10mm}

This Supplementary material consists of three sections. In Sec. 1, we derive the correction to the self-energy of an electron in the PC medium
given by Eq. (7) in the main text. Section 2 is devoted to deriving and explaining the correction to the electron mass in one-dimensional PCs.
In Sec. 3, we show that the PC correction to the electron mass is free from ultraviolet divergences.

\section*{S.1 Correction to the self-energy of an electron in the PC medium}\label{S1}
In this section we present a derivation of Eq.~(7) of the main text. We start from Eq.~(6)
\begin{equation}\label{Seqn11}
\begin{split}
{{\Delta }{E_{em}^{pc}}}(\textbf{p}) = {\sum\limits_{{\bf{p}}'}\sum\limits_{{\bf{k}},\;{{n}}} {\frac{{\left\langle {\bf{p}}
\right|\widehat{H}_I^{pc}\left| {{\bf{p}}';{\bf{k}},n} \right\rangle \left\langle {{\bf{p}}';{\bf{k}},n} \right|\widehat{H}_I^{pc}\left| {\bf{p}} \right\rangle
}}{{\frac{{{{\bf{p}}^2}}}{{2{m_e}}} - \frac{{{\bf{p}}{'^2}}}{{2{m_e}}} - {\omega _{{\bf{k}}n}}}}} } - \\
- {\sum\limits_{{\bf{p}}'}\sum\limits_{{\bf{k}},\;{\lambda }} \frac{{\left\langle {\bf{p}} \right|\widehat{H}_I^{}\left|
{{\bf{p}}';{\bf{k}},{\bm{\varepsilon}_\lambda }} \right\rangle \left\langle {{\bf{p}}';{\bf{k}},{\bm{\varepsilon}_\lambda }}
\right|\widehat{H}_I^{}\left|
{\bf{p}} \right\rangle }}{{\frac{{{{\bf{p}}^2}}}{{2{m_e}}} - \frac{{{\bf{p}}{'^2}}}{{2{m_e}}} - \left| {\bf{k}} \right|}}}.
\end{split}
\end{equation}
The matrix element $\left\langle {{\bf{p}}';{\bf{k}},n} \right|\widehat{H}_I^{pc}\left| {\bf{p}} \right\rangle $ of the interaction Hamiltonian
$\widehat{H}_I^{pc}=-\frac{e}{m_e}{\widehat{\textbf{p}}} \cdot \widehat {\bf{A}}_{pc}({\bf{r}})$ with
\begin{equation}\label{eqn11b}
\begin{split}
{\widehat{\bf{A}}_{pc}}({\bf{r}},t) = \sum\limits_{{\bf{k}},\;n} {\left[ {{{\bf{A}}_{{\bf{k}}n}}({\bf{r}}){{\hat a}_{{\bf{k}}n}}{e^{ -
i{\omega _{{\bf{k}}n}}t}} + {\bf{A}}_{{\bf{k}}n}^*({\bf{r}})\hat a_{{\bf{k}}n}^\dag {e^{i{\omega _{{\bf{k}}n}}t}}} \right]},
\end{split}
\end{equation}
where ${{\bf{A}}_{{\bf{k}}n}}({\bf{r}}) = {1 \mathord{\left/{\vphantom {1 {\sqrt {V{\omega _{{\bf{k}}n}}} }}} \right.
 \kern-\nulldelimiterspace} {\sqrt {V{\omega _{{\bf{k}}n}}} }} {{\bf{E}}_{{\bf{k}}n}}({\bf{r}})$ with $\textbf{E}_{\textbf{k}n}(\textbf{r})$
 being the Bloch eigenfunctions ${\bf{E}}_{{\bf{k}}n} ({\bf{r}}) =
\sum_{\bf{G}}{{\bf{E}}_{{\bf{k}}n} ({\bf{G}})e^{i\left( {{\bf{k}} + {\bf{G}}} \right) \cdot {\bf{r}}} }$ can be represented in the form
\begin{equation}\label{Seqn12}
\begin{split}
\left\langle {{\bf{p}}';{\bf{k}},n} \right|\widehat{H}_I^{pc}\left| {\bf{p}} \right\rangle  = \;\;\;\;\;\;\;\;\;\;\;\;\;\;\;\;\\
= - \frac{e}{m_e}\int {{d^3}r} \Psi
_{{\bf{p}}'}^*({\bf{r}})( - i{\nabla _{\bf{r}}}{{\bf{A}}_{{\bf{k}}n}^*}({\bf{r}}))\Psi _{\bf{p}}^{}({\bf{r}}) = \\
= \frac{e}{{m_e{V^{3/2}}\sqrt {{\omega _{{\bf{k}}n}}} }}\int {{d^3}r} {e^{ - i{\bf{p}}'{\bf{r}}}}(i{\nabla
_{\bf{r}}}{{\bf{E}}_{{\bf{k}}n}^*}({\bf{r}})){e^{i{\bf{pr}}}}
\end{split}
\end{equation}
with ${\Psi _{\bf{p}}}({\bf{r}})$ being the normalized wave function of the electron state ${\Psi _{\bf{p}}}({\bf{r}}) = \left\langle
{{\bf{r}}}\mathrel{\left | {\vphantom {{\bf{r}} {\bf{p}}}}\right. \kern-\nulldelimiterspace}{{\bf{p}}} \right\rangle$.
Here we have taken into account that ${\Psi _{\bf{p}}} = {e^{i{\bf{pr}}}}/\sqrt V $ for ${\bf{r}} \in V$ and ${\Psi _{\bf{p}}} = 0$ for
${\bf{r}} \notin V$. For $\left\langle {{\bf{p}}';{\bf{k}},n} \right|\widehat{H}_I^{pc}\left| {\bf{p}} \right\rangle$ we get
\begin{equation}\label{Seqn13}
\left\langle {{\bf{p}}';{\bf{k}},n} \right|\widehat{H}_I^{pc}\left| {\bf{p}} \right\rangle  =  - \frac{e}{m_e}\frac{1}{{\sqrt {V{\omega _{{\bf{k}}n}}}
}}\sum\limits_{\bf{G}} {{\bf{p}} \cdot {{\bf{E}}_{{\bf{k}}n}^{*}}({\bf{G}})} {\delta _{{\bf{p}},{\bf{q}}}}
\end{equation}
with ${\bf{q}} = {\bf{p'}} + {\bf{k}} + {\bf{G}}$. In the same way, for $\left\langle {\bf{p}} \right|\widehat{H}_I^{pc}\left| {{\bf{p}}';{\bf{k}},n}
\right\rangle$ we find
\begin{equation}\label{Seqn14}
\left\langle {\bf{p}} \right|\widehat{H}_I^{pc}\left| {{\bf{p}}';{\bf{k}},n} \right\rangle  =  - \frac{e}{m_e}\frac{1}{{\sqrt {V{\omega _{{\bf{k}}n}}}}}
\sum\limits_{\bf{G}} {{\bf{p}} \cdot {\bf{E}}_{{\bf{k}}n}({\bf{G}})} {\delta _{{\bf{p}},{\bf{q}}}}.
\end{equation}
Correspondingly, for the matrix elements of the interaction Hamiltonian in the free space, we have
\begin{equation}\label{Seqn15}
\left\langle {{\bf{p}}';{\bf{k}}{\bf{,}}{\bm{\varepsilon}_\lambda }} \right|\widehat{H}_I^{}\left| {\bf{p}} \right\rangle  =  -
\frac{e}{m_e}\frac{1}{{\sqrt {2V\left| {\bf{k}} \right|} }}\sum\limits_\lambda  {{\bf{p}} \cdot {\bm{\varepsilon}_\lambda }({\bf{k}})} {\delta
_{{\bf{p}},{\bf{q}}}},
\end{equation}
\begin{equation}\label{Seqn16}
\left\langle {{\bf{p}}} \right|\widehat{H}_I^{}\left| {{\bf{p}}';{\bf{k}}{\bf{,}}{\bm{\varepsilon}_\lambda }} \right\rangle  =  -
\frac{e}{m_e}\frac{1}{{\sqrt {2V\left| {\bf{k}} \right|} }}\sum\limits_\lambda  {{\bf{p}} \cdot {\bm{\varepsilon}_\lambda }({\bf{k}})} {\delta
_{{\bf{p}},{\bf{q}}}}
\end{equation}
with ${\bf{q}} = {\bf{p'}} + {\bf{k}}$. Substituting these matrix elements of interaction Hamiltonians $\widehat{H}_I^{pc}$ and $\widehat{H}_I$ into
Eq.~(\ref{Seqn11}), and replacing the discrete sums by integrals $\sum\nolimits_{{\bf{k}}, \;n}  \, \to \tfrac{V}{{{{(2\pi
)}^3}}}\sum\nolimits_n  \,\smallint {d^3}{\bf{k}}$ and $\sum\nolimits_{\bf{k}}  \, \to \tfrac{V}{{{{(2\pi )}^3}}}\smallint {d^3}{\bf{k}}$ we
get
\begin{equation}\label{Seqn17}
\begin{split}
\Delta E_{em}^{pc}(\widehat {\bf{p}}) =  - \frac{\alpha }{{2m_e^2{\pi ^2}}}\left( {\sum\limits_\lambda  {\int\limits_{} {\frac{{{d^3}{\bf{k}}}}
{{2\left| {\bf{k}} \right|}}\frac{{{{\left| {\widehat {\bf{p}} \cdot {\bm{\varepsilon} _\lambda }({\bf{k}})} \right|}^2}}}{{\frac{{{{\widehat {\bf{p}}}^2}}}
{{2{m_e}}} - \frac{{{{\left( {\widehat {\bf{p}} - {\bf{k}}} \right)}^2}}}{{2{m_e}}} - \left| {\bf{k}} \right|}}} } } \right. - \\
\left. { - \sum\limits_{n,\;{\bf{G}}} {\int\limits_{FBZ} {\frac{{{d^3}{\bf{k}}}}{{{\omega _{{\bf{k}}n}}}}\frac{{{{\left| {\widehat {\bf{p}}
\cdot {{\bf{E}}_{{\bf{k}}n}}\left( {\bf{G}} \right)} \right|}^2}}}{{\frac{{{{\widehat {\bf{p}}}^2}}}{{2{m_e}}} -
\frac{{{{\left( {\widehat {\bf{p}} - {\bf{k}} - {\bf{G}}} \right)}^2}}}{{2{m_e}}} - {\omega _{{\bf{k}}n}}}}} } } \right).
\end{split}
\end{equation}
\section*{S.2 Correction to the electron mass in one-dimensional PCs}\label{S2}
The PC correction to the electron mass can be represented in the form
\begin{equation}\label{Seqn21}
{\delta m}_{pc}\left(\widehat{\bf{I}}_{\bf{p}} \right) = \Delta m_{em}^{pc}\left(\widehat{\bf{I}}_{\bf{p}} \right) - \Delta m_{em},
\end{equation}
where $\Delta m_{em}^{pc}\left(\widehat{\bf{I}}_{\bf{p}} \right)$ and $\Delta m_{em}$ are the total electromagnetic mass in PC and in vacuum. They are both divergent in standard
QED, but this PC correction to the electromagnetic mass is finite and could be calculated. The PC correction to the electron mass is given
by equation (Eq.~(9) in the main text)\cite{Gainutdinov2012}
\begin{equation}\label{eqn211}
\begin{split}
{\delta m}_{pc}\left(\widehat{\bf{I}}_{\bf{p}} \right) = \frac{\alpha }{{\pi ^2 }}\left[ {\sum\limits_{n,\;{\bf{G}}} {\int {{k_\rho }d{k_\rho }} \int\limits_{FBZ}^{} {d{k_z}
\int {d\varphi } } }\cdot }\right. \\
\cdot\frac{{\left| {{\widehat{\bf{I}}_{\bf{p}}}\cdot {\bf{E}}_{{\bf{k}}n} ({\bf{G}})} \right|^2 }}{{\omega _{{\bf{k}}n}^2}}
- \left.{\int {\frac{{d^3 {\bf{k}}}}{{2{\bf{k}}^2 }}} \sum\limits_{\lambda  = 1}^2  \,\mathop {\left| { {\widehat{\bf{I}}_{\bf{p}}}\cdot
 \bm{\varepsilon} _\lambda  ({\bf{k}})} \right|}\nolimits^2 } \right].
\end{split}
\end{equation}
Representing the electromagnetic field in 1D PC in the Bloch form (Eq.~(13) in the main text), the PC correction can be rewritten as
\begin{equation}\label{Seqn22}
{\delta m}_{pc}\left(\widehat{\bf{I}}_{\bf{p}} \right) = \sum\limits_{i=1}^3\Delta {m_i}\left(\widehat{\bf{I}}_{\bf{p}} \right) - \Delta m_{em}
\end{equation}
with
\begin{equation}\label{Seqn23}
\begin{split}
\Delta {m_1}\left(\widehat{\bf{I}}_{\bf{p}} \right) = \frac{\alpha }{{{\pi ^2}}}\sum\limits_{n,\ G} {{\int {{k_\rho }d{k_\rho }} \int\limits_{FBZ}^{} {d{k_z}\int {d\varphi } } }}\cdot \\
\cdot{\frac{{{{\left( {f_{{\bf{p}} \cdot {\bm{\varepsilon}}_{{1}} }^{}E_{{\bf{k}}n1}(G)} \right)}^2}}}{{\omega _{{\bf{k}}n1}^2}}  },
\end{split}
\end{equation}
\begin{eqnarray}\label{Seqn24}
\begin{split}
\Delta {m_2}\left(\widehat{\bf{I}}_{\bf{p}} \right) = \frac{\alpha }{{{\pi ^2}}}\sum\limits_{n,\;G}^{} {{\int {{k_\rho }d{k_\rho }} \int\limits_{FBZ}^{} {d{k_z}\int {d\varphi } } }} \cdot \\
\cdot {\frac{{2 {{f_{{\bf{p}} \cdot {\bm{\varepsilon}}_{{1}}}}{E_{{\bf{k}}n1}}(G)} }}{{\omega _{{\bf{k}}n1}}}}\frac{{ {{f_{{\bf{p}} \cdot {\bm{\varepsilon}}_{{2}}}}{E_{{\bf{k}}n2}}(G)} }}{{\omega _{{\bf{k}}n2}}},
\end{split}
\end{eqnarray}
\begin{equation}\label{Seqn25}
\begin{split}
\Delta {m_3}\left(\widehat{\bf{I}}_{\bf{p}} \right) = \frac{\alpha }{{{\pi ^2}}}\sum\limits_{n,\ G} {{\int {{k_\rho }d{k_\rho }} \int\limits_{FBZ}^{} {d{k_z}\int {d\varphi }}}} \cdot\\
\cdot{\frac{{{{\left( {f_{{\bf{p}} \cdot {\bm{\varepsilon}}_{{2}} }^{}E_{{\bf{k}}n2}(G)} \right)}^2}}}{{\omega _{{\bf{k}}n2}^2}} },
\end{split}
\end{equation}
where $f_{{\bf{p}} \cdot {\bm{\varepsilon}}}^{} = \sin {\Theta _{\bm{\varepsilon}}}\sin \Theta \cos (\Phi  - {\Phi _{\bm{\varepsilon}}}) +
\cos {\Theta _{\bm{\varepsilon}}}\cos \Theta$ is the scalar product between the unit vector of electron's momentum with angular coordinates
$\left( {\Theta ,\Phi } \right)$ and ${\bm{\varepsilon}}_{{1}}(\bf{k_G}) $ and ${\bm{\varepsilon}}_{{2}}(\bf{k_G})$ are the field unit vectors
with angular coordinates $\left( {{\Theta _{\bm{\varepsilon}}},{\Phi _{\bm{\varepsilon}}}} \right)$ and ${\bf{k_G}} = {\bf{k}} +
G{{\bf{e}}_z}$ in the case of 1D PC. Taking the integrals in Eqs.~(\ref{Seqn23}--\ref{Seqn25}) over the azimuthal angle $\varphi_k$ in the
cylindrical coordinate system, we get
\begin{equation}\label{Seqn26}
\begin{split}
\Delta {m_1}\left(\widehat{\bf{I}}_{\bf{p}} \right) = \frac{{{\alpha }}}{{{\pi }}}\sum\limits_{n,\ G}^{} {\int {{k_\rho }d{k_\rho }} \int\limits_{FBZ}^{} {d{k_z}
\frac{{{{\left| {{E_{{\bf{k}}n1}}(G)} \right|}^2}}}{{{{\omega }}_{{\bf{k}}n1}^2}} \cdot } }\\
\cdot \left( {\frac{{(k_z + G)^2}}{{k_\rho ^2 + (k_z + G)^2}}{{\sin }^2}\Theta  + 2\frac{{k_\rho ^2}}{{k_\rho ^2 + (k_z + G)^2}}
{{\cos }^2}\Theta } \right),
\end{split}
\end{equation}
\begin{equation}\label{Seqn27}
\Delta {m_2}\left(\widehat{\bf{I}}_{\bf{p}} \right) = 0,
\end{equation}

\begin{equation}\label{Seqn28}
\Delta {m_3}\left(\widehat{\bf{I}}_{\bf{p}} \right) = \frac{{{\alpha }}}{{{\pi }}}\sum\limits_{n,\ G}^{} {\int {{k_\rho }d{k_\rho }} \int\limits_{FBZ}^{} {d{k_z}} \frac{{{{\left|
{{E_{{\bf{k}}n2}}(G)} \right|}^2}}}{{{{\omega }}_{{\bf{k}}n2}^2}}{{\sin }^2}\Theta },
\end{equation}
where $k_\rho, k_z + G$ are radial component and $z$-component of the wave vector $\bf{k}$ in cylindrical coordinate system. The
electromagnetic mass of an electron in vacuum takes the form $\frac{{4{{\alpha }}}}{{3{{\pi }}}}{\int {dk}}$. Then the PC correction to the
electron mass Eq.~(\ref{Seqn22}) in 1D PC is equal:
\begin{equation}\label{Seqn29}
\begin{split}
\delta {m_{pc}}\left( {{{\widehat {\bf{I}}}_{\bf{p}}}} \right) = \frac{\alpha }{\pi }\sum\limits_{n,\;G}^{} {\int {{k_\rho }d{k_\rho }}
\int\limits_{FBZ}^{} {d{k_z}\left[ {\frac{{{{\left| {{E_{{\bf{k}}n1}}(G)} \right|}^2}}}{{\omega _{{\bf{k}}n1}^2}} \cdot } \right.} } \\
\left. { \cdot \frac{{\left( {{{({k_z} + G)}^2}{{\sin }^2}\Theta  + 2k_\rho ^2{{\cos }^2}\Theta } \right)}}{{k_\rho ^2 + {{({k_z} + G)}^2}}}
+ \frac{{{{\left| {{E_{{\bf{k}}n2}}(G)} \right|}^2}}}{{\omega _{{\bf{k}}n2}^2}} \cdot {{\sin }^2}\Theta } \right] - \\
 - \frac{{4\alpha }}{{3\pi }}\int {dk}.
\end{split}
\end{equation}
Thus the operator ${\delta m}_{pc}\left(\widehat{\bf{I}}_{\bf{p}} \right)$ can be presented in the form
\begin{equation}\label{Seqn210}
{\delta {m}}_{pc}\left(\widehat{\bf{I}}_{\bf{p}} \right) = A + \left( {{{\widehat {\bf{I}}}_{\bf{p}}} \cdot {{\widehat {\bf{I}}}_{pc}}} \right)^{2}B,
\end{equation}
where ${\widehat {\bf{I}}_{\bf{{p}}}} = \frac{\widehat{\bf{p}}}{\left| \widehat{\bf{p}} \right|}$  being the direction of the electron momentum,
${{{\widehat {\bf{I}}}_{pc}}}$ is the unit vector of the 1D PC crystal axis that coincides with vector ${\bf{e}}_z$, $\left( {{{\widehat
{\bf{I}}}_{\bf{p}}} \cdot {{\widehat {\bf{I}}}_{pc}}} \right) = \cos \Theta$, and
\begin{equation}
\begin{split}
A = \frac{{{\alpha }}}{{{\pi }}}\sum\limits_{n,\;G}^{} {\int {{k_\rho }d{k_\rho }} \int\limits_{FBZ}^{} {d{k_z}\left( {\frac{{{{\left|
{E_{{\bf{k}}n1}^{}\left( G \right)} \right|}^2}}}{{{{\omega }}_{{\bf{k}}n1}^2}}\frac{{(k_z + G)^2}}{{k_\rho ^2 +(k_z + G)^2}}}\right.}} \\
\left. { + \frac{{{{\left| {E_{{\bf{k}}n2}^{}\left( G \right)} \right|}^2}}}{{{{\omega }}_{{\bf{k}}n2}^2}}} \right)-\frac{{4{{\alpha
}}}}{{3{{\pi }}}}{\int {dk}},\notag \\
\end{split}
\end{equation}
\begin{equation}
\begin{split}
B = \frac{{{\alpha }}}{{{\pi }}}\sum\limits_{n,\;G}^{} {\int {{k_\rho }d{k_\rho }} \int\limits_{FBZ}^{} {d{k_z}\left( {\frac{{{{\left|
{E_{{\bf{k}}n1}^{}\left( G \right)} \right|}^2}}}{{{{\omega }}_{{\bf{k}}n1}^2}}} \cdot\right.} }\\
\cdot\frac{{2k_\rho ^2 - (k_z + G)^2}}{{k_\rho ^2 + (k_z + G)^2}} \left. { - \frac{{{{\left| {E_{{\bf{k}}n2}^{}\left( G \right)} \right|}^2}}}
{{{{\omega }}_{{\bf{k}}n2}^2}}} \right).\notag \\
\end{split}
\end{equation}

\section*{S.3 Finiteness of the PC correction to the electron mass}\label{S3}
In the investigation of the convergence of the integrals in Eq.~(\ref{Seqn29}) it is natural to consider the PC correction to the
electromagnetic mass in the high-energy limit of photons \cite{Schweber2011}
\begin{equation}\label{Seqn33}
\begin{split}
{\delta m}_{pc}\left(\widehat{\bf{I}}_{\bf{p}} \right) = \frac{\alpha }{{{\pi ^2}}}\left[ {\mathop {\sum\limits_{n,\;G} {\int\limits_{} {{k_\rho }d{k_\rho }} } }\limits^{{\omega
_{{\bf{k}}n}} > {\omega ({{\bf{k}}_0})} } \int\limits_{FBZ}^{} {d{k_z}\int\limits_0^{2\pi } {d{k_\varphi }}  } } \right. \cdot \\
\cdot \sum\limits_{\lambda  = 1}^2 {\frac{{{{\left| {{E_{{\bf{k}}n\lambda }}(G)} \right|}^2}}}{{\omega _{{\bf{k}}n\lambda }^2}}} {\left|
{{\widehat{\bf{I}}_{\bf{p}}} \cdot {\bm{\varepsilon} _\lambda }({{\bf{k}}_{\bf{G}}})} \right|^2} - \left. {\int\limits_{\left| {\bf{k}} \right| >
k_0 }^{\infty} {\frac{{{d^3}{\bf{k}}}}{{3{{\bf{k}}^2}}}} } \right].
\end{split}
\end{equation}
Here the value $k_0$ should be chosen to be much larger than the width $|{{\bf{b}}_z}|$ of FBZ. In the limit $\bf{k} \to \infty$ and, as a consequence,
$\omega({\bf{k}}) \to \infty$, the 1D PC medium is considered as a free space with effective
refractive index ${\tilde{n}}_{eff}({\bf{k}})$ defined by equation \cite{Skorobogatiy2009}
\begin{equation}\label{Seqn34}
{\tilde{n}}_{eff}({\bf{k}}) = ({n_h({\bf{k}})d_h + n_l d_l})/({d_h + d_l}),
\end{equation}
where $d_h$ and $d_l$ are the thicknesses corresponding to the layers of the 1D PC with higher (\textit{h}) and lower (\textit{l}) refractive
index $n_h({\bf{k}})$ and $n_l = 1$ (air voids). In this limit it is more convenient to make use the extended zone scheme, where the summation
on the band index $n$, and the integration in the FBZ are transformed into the integral over all wave vectors in space satisfying the
condition $|{\bf{k}}| > k_0$
\begin{equation}\label{Seqn35}
\begin{split}
{\delta m}_{pc}\left(\widehat{\bf{I}}_{\bf{p}} \right) = \frac{\alpha }{{{\pi ^2}}}\left[ {\int\limits_{\left| {\bf{k}} \right| > k_0 }^{\infty} {{k_\rho }d{k_\rho }}
\int\limits_{\left| {\bf{k}} \right| > k_0 }^{\infty} {d{k_z}\int\limits_0^{2\pi } {d{k_\varphi }}  } } \cdot \right. \\
\left. { \cdot \sum\limits_{\lambda  = 1}^2 {\frac{{{{\left| {{E_\lambda }({\bf{k}})} \right|}^2}}}{\omega_{\lambda}^{2} ({\bf{k}})}{{\left|
{{\widehat{\bf{I}}_{\bf{p}}} \cdot {\bm{\varepsilon} _\lambda }({{\bf{k}}})} \right|}^2}}  - \int\limits_{\left| {\bf{k}} \right| > k_0 }^{\infty}
{\frac{{{d^3}{\bf{k}}}}{{3{{\bf{k}}^2}}}} } \right].
\end{split}
\end{equation}
Then dispersion relations $\omega_{\lambda} ({\bf{k}})$ are symmetrical quasi-continuous functions of ${\bf{k}}$ everywhere in reciprocal
space \cite{Ashcroft1976}
\begin{equation}\label{Seqn36}
\omega_{\lambda} ({\bf{k}})=\frac{|{\bf{k}}|}{{\tilde{n}}_{eff}({\bf{k}})} \left(1 + \frac{{{\tilde{n}}_{eff}({\bf{k}})}\Delta
\omega_{\lambda}({\bf{k}})}{|{\bf{k}}|} \right) \\
\end{equation}
with $\Delta \omega_{\lambda}({\bf{k}})$ being the PC correction to the dispersion relation $\frac{|{\bf{k}}|}{{\tilde{n}}_{eff}({\bf{k}})}$
in isotropic medium and
\begin{equation}\label{Seqn37}
|E_{\lambda}({\bf{k}})|^2 = \frac{1}{2} + \Delta E_{\lambda}({\bf{k}}). \\
\end{equation}
These eigenfunctions of Maxwell's equations with corresponding eigenvalues $\omega_{\lambda} ({\bf{k}})$ are equal to $\frac{1}{2}$ because
$\Delta E_{\lambda}({\bf{k}})$ is proportional to $\Delta \omega_{\lambda}({\bf{k}})$.
The correction $\Delta \omega_{\lambda}({\bf{k}})$ is limited by the function \cite{Skorobogatiy2009}
\begin{equation}\label{Seqn38}
\Delta \omega_{\lambda}({\bf{k}}) \leq \frac{n_h({\bf{k}}) - 1}{n_h({\bf{k}}) + 1}\omega. \\
\end{equation}
According to Sellmeier equation \cite{Ghosh1997}, ${{n}}_{h} ({\textbf{k}})$ can be represented as a power series
\begin{equation}\label{Seqn39}
{{n}}_{h} ({\textbf{k}}) = 1 + \frac{C_1}{{\bf{k}}^2} + \frac{C_2}{{\bf{k}}^4} + ... \\
\end{equation}
where $C_{1,2}$ are some experimentally determined parameters. Under the condition of high-energy photons propagating in 1D PC the corrections
$\Delta \omega_{\lambda}({\bf{k}})$ in dispersion relations are vanished because $n_h({\bf{k}}) - 1\rightarrow 0$ as $C_1/{\textbf{k}}^2$ and
eigenfrequencies $\omega_{\lambda}({\bf{k}})$ tend to each other. Then the PC contribution to the electromagnetic mass (\ref{Seqn35}) is represented
in the form
\begin{equation}
\label{Seqn310}
\begin{split}
\delta m_{pc} = \frac{\alpha }{{3{\pi ^2}}}\int\limits_{|{\bf{k}}| > k_0 }^\infty  {{d^3}{\bf{k}}{\frac{{\tilde n_{eff}^2({\bf{k}}) -
1}}{{{{\bf{k}}^2}}} + O\left( {\frac{{C_1}}{{{{k_0}}}}} \right) + O\left( {\frac{{{\bf{b}}_z^2}}{{{{k_0}}}}} \right)}}.
\end{split}
\end{equation}
From Eqs.~(\ref{Seqn34}) and (\ref{Seqn39}) it follows that
\begin{equation}\label{Seqn311}
\begin{split}
\delta {m_{pc}} = \frac{\alpha }{{6{\pi ^2}}}\int\limits_{|{\bf{k}}| > k_0 }^\infty  {\frac{{{d^3}{\bf{k}}}}{{{{\bf{k}}^4}}} {
\frac{C_1 d_h}{d_h + d_l} + O\left( {\frac{{C_1}}{{{{k_0}}}}} \right) + O\left( {\frac{{{\bf{b}}_z^2}}{{{{k_0}}}}} \right)} }.
\end{split}
\end{equation}
Thus the PC correction to the electromagnetic mass is free from the ultraviolet divergences and hence is an observable.

\bibliography{cas-refs}

\end{document}